\documentclass[11pt,a4paper]{article}
\usepackage{amssymb,fullpage,amsmath}
\usepackage{lineno}
\usepackage{amsfonts,amsbsy}
\usepackage{graphicx}
\usepackage{color}
\usepackage{mathrsfs}

\usepackage[authoryear]{natbib}
\setcitestyle{aysep={}}

\newcommand*\patchAmsMathEnvironmentForLineno[1]{%
  \expandafter\let\csname old#1\expandafter\endcsname\csname #1\endcsname
  \expandafter\let\csname oldend#1\expandafter\endcsname\csname end#1\endcsname
  \renewenvironment{#1}%
     {\linenomath\csname old#1\endcsname}%
     {\csname oldend#1\endcsname\endlinenomath}}%
\newcommand*\patchBothAmsMathEnvironmentsForLineno[1]{%
  \patchAmsMathEnvironmentForLineno{#1}%
  \patchAmsMathEnvironmentForLineno{#1*}}%
\AtBeginDocument{%
\patchBothAmsMathEnvironmentsForLineno{equation}%
\patchBothAmsMathEnvironmentsForLineno{align}%
\patchBothAmsMathEnvironmentsForLineno{flalign}%
\patchBothAmsMathEnvironmentsForLineno{alignat}%
\patchBothAmsMathEnvironmentsForLineno{gather}%
\patchBothAmsMathEnvironmentsForLineno{multline}%
}

\let\ssection=\section
\renewcommand{\section}{\setcounter{equation}{0}\ssection}

\usepackage{amssymb}
\usepackage{amsmath}
\usepackage{amsfonts}
\usepackage{tikz}  
\usetikzlibrary{calc,intersections}

\usepackage{color}

\numberwithin{equation}{section}

\renewcommand{\d}{d} 
\newcommand{\ip}{\lrcorner}

\newcommand{\divv}{\mathop{\mathrm{div}}\nolimits}

\newcommand{\Ac}{\mathcal{A}}
\newcommand{\Bc}{\mathcal{B}}
\newcommand{\lie}{\mathcal{L}}
\newcommand{\Mc}{\mathcal{M}} 
\newcommand{\Nc}{\mathcal{N}} 
 
\newcommand{\Vc}{\mathcal{V}}

\newcommand{\fv}{\boldsymbol{f}}
\newcommand{\nv}{\boldsymbol{n}}
\newcommand{\uv}{\boldsymbol{u}}
\newcommand{\xv}{\boldsymbol{x}}

\newcommand{\sigmav}{\boldsymbol{\sigma}}

\newcommand{\cd}{\mathfrak{d}}

\newcommand{\eps}{\varepsilon}

\newcommand{\dwedge}{\overset{\cdot}{\wedge}}

\newcommand{\dt}[2]{\frac{\mathrm{d} #1}{\mathrm{d} #2}}

\newcommand{\byparts}{\; \;\simeq\; \;} 

\definecolor{airforceblue}{rgb}{0.36, 0.54, 0.66}

\title{A geometric look at momentum flux and stress in fluid mechanics}

\author{Andrew D. Gilbert$^1$ and Jacques Vanneste$^2$\footnote{Corresponding author: J.Vanneste@ed.ac.uk}}
\date{
\normalsize{$^{1}$Department of Mathematics, College of Engineering, Mathematics and Physical Sciences, University of Exeter, EX4 4QF, UK \\
$^{2}$School of Mathematics and Maxwell Institute for Mathematical Sciences, University of Edinburgh, Edinburgh EH9 3FD, UK}}

\begin{document}

\maketitle

\begin{abstract}
We develop a geometric formulation of fluid dynamics, valid on arbitrary Riemannian manifolds, that regards the momentum-flux and stress tensors as 1-form-valued 2-forms, and their divergence as a covariant exterior derivative. We review the necessary tools of differential geometry and obtain the corresponding coordinate-free form of the equations of motion for a variety of inviscid fluid models -- compressible and incompressible Euler equations, Lagrangian-averaged Euler-$\alpha$ equations, magneto\-hydrodynamics and shallow-water models -- using a variational derivation which automatically yields a symmetric momentum flux. We also consider dissipative effects and discuss the geometric form of the Navier--Stokes equations for viscous fluids and of the Oldroyd-B model for visco-elastic fluids.
\end{abstract}

\section{Introduction}

The equations of fluid dynamics are traditionally presented in coordinate forms, typically using Cartesian coordinates. There are advantages, however, in geometrically intrinsic formulations which highlight the underlying structure of the equations, apply to arbitrary manifolds and, when the need arises, are readily translated into whatever coordinate system is convenient. The most straightforward geometric formulations rely on the advective form of the momentum equation, with the advective derivative expressed in terms of Lie or covariant derivatives \citep{ArKh98,Fr,Sc80,HoScSt09}.
 One benefit of the Lie-derivative form is that the metric appears only undifferentiated, in the relationship between advected momentum and  advecting velocity.
It is in this form  that the Euler equations and more general inviscid fluid models emerge from variational arguments as so-called Euler--Poincar\'e systems \citep{Ar66,Sa88,Mo98,ArKh98,HoMaRa,We18}. 

An alternative to the advective form of the momentum equation is the conservation form, in which the material advection term is replaced by the divergence of the momentum flux. The conservation form is particularly useful for its close relationship to the global conservation law of (volume-integrated) momentum, when such a law holds. It is also useful in the context of Reynolds averaging and its extensions, where the effect of unresolved fluctuations naturally emerges as the divergence of the Reynolds stress, the fluctuation-averaged momentum flux.
In Euclidean space and for the Euler equations, it is straightforward to switch between the two forms and to derive global conservation laws for momentum in each spatial direction. It is less straightforward on other manifolds, where global momentum conservation laws exist only in the presence of spatial symmetries, and for fluid models more complicated than the Euler equations. This points to the benefits of formulating fluid models in conservation form in a geometrically intrinsic way. This is the first objective of this paper. The second is to discuss the geometric nature of the Cauchy stress tensor (associated with pressure and irreversible effects) and of its divergence, noting that the momentum-flux and stress tensors enter the equations of fluid mechanics on a similar footing.

A first question concerns the geometric interpretation of these tensors. We follow \citet{KaArToYaMaDe} and regard them fundamentally as 1-form valued 2-forms, related to the more familiar twice contravariant tensors through operations involving the metric. In this formulation, the divergence of the momentum flux and stress tensors becomes  
the covariant exterior derivative of the associated 1-form valued 2-forms. Defining and manipulating these objects requires some differential-geometric machinery which we introduce in \S\ref{secmachinery}. The interpretation of momentum flux and stress as 1-form valued 2-forms (or their close relatives, namely vector valued 2-forms) is advocated by \citet{Fr} who points to its origin in the work of E. Cartan and Brillouin. It has both conceptual and practical benefits. First, 1-form valued 2-forms arise naturally when the stress is regarded as a force, to be paired with a velocity field and integrated over a surface to obtain a rate of work. Second, it enables a simple coordinate-free formulation that makes minimal use of the metric and associated connection. The computations, of the covariant exterior derivative in particular, are then straightforward when carried out at the level of differential forms rather than coordinates. We illustrate this by computation in spherical geometry in appendix \ref{app:sphere} (see also \citet{Fr} for similar computations using vector valued forms in the context of solid mechanics). Third, the formulation proves useful for the derivation of momentum-conserving discretisations of the Navier--Stokes equations \citep{ToHuGe,Ge}.

We  consider the derivation of fluid equations in their conservation form, starting with the Euler equations for compressible perfect fluids in \S\ref{secidealflow}. We follow two routes. The first takes the Euler equations in their advective form as starting point, and uses a relation between Lie derivative and covariant exterior derivative to deduce the conservation form. The second relies on a variational formulation of the Euler equations: we show that the stationarity of the relevant action functional, when combined with an infinitesimal condition for the covariance of the action (that is, for its invariance with respect to arbitrary changes of variables),  leads  directly to the Euler equations in their conservation form. The variational route has the benefit of being systematic and of automatically yielding the momentum flux as a symmetric 1-form valued 2-form. We follow this route to derive the conservation form of further inviscid fluid models: the incompressible Euler equations in \S\ref{secincomp}, the Lagrangian-averaged Euler $\alpha$-model in \S\ref{secalpha} and the magnetohydrodynamics (MHD) equations in \S\ref{secmhd}. Analogous derivations for the shallow-water model and its MHD extension are sketched in Appendix \ref{apshallow}. We emphasise that, for models such as the Euler-$\alpha$ model, the form of the momentum flux does not follow readily from the advective form of the equations, even in Euclidean geometry, making the variational derivation valuable.

In \S\ref{secviscous} we examine the interpretation of the Cauchy stress tensor as a 1-form valued 2-form for Newtonian and viscoelastic fluids. In the Newtonian case, we give an expression for the viscous stress tensor in terms of the Lie derivative of the metric tensor along the fluid flow, and we emphasise the significance of this derivative as a natural measure of the rate of deformation of the fluid. In the conservation form of the Navier--Stokes equations which emerges by taking a covariant exterior derivative, the viscous term involves the Ricci Laplacian of the momentum. This Laplacian differs from both the Laplace-de Rham operator and the rough Laplacian by terms proportional to the Ricci tensor. Its appearance is consistent with physical arguments \citep{GiRiTh}. For viscoelastic fluids, we discuss models whose constitutive laws involve the transport of the stress tensors and sketch a geometric derivation of the constitutive law of one standard representative of this class, the Oldroyd-B model. The formulation in terms of 1-form (or vector) valued 2-forms sheds light on the reasons underlying the appearance of a particular type of material derivative of the stress tensor (the upper-convected derivative in this instance). 

Many of the concepts and techniques presented in this paper are standard and discussed in existing literature on differential geometry and on geometric mechanics. Their use in fluid dynamics is, however, not well established. By introducing them in the context of familiar fluid models we aim to promote their adoption more broadly in fluid dynamics and its applications.

\section{Machinery}
\label{secmachinery}

We will be using techniques of differential geometry and work on a smooth, orientable Riemannian manifold $\Mc$, with or without a boundary $\partial \Mc$. 
We take $\Mc$ to be three-dimensional, although formulae and arguments are easily modified for the two-dimensional case. To avoid  unnecessary complications we assume $\Mc$ has a straightforward topology, so that all curves and surfaces in $\Mc$ may be contracted to a point. The manifold is  equipped with a metric $g$ and we also need the compatible volume form $\mu$ and covariant derivative $\nabla$. We assume that the reader is familiar with the fundamental constructions of differential geometry including vectors, $p$-forms, the interior product $\ip$, the Lie derivative $\lie$, the exterior derivative $d$, the Hodge star operator $\star$, and the musical raising and lowering operators $\sharp$ and $\flat$ (see for example \citet{Fr, Sc80, HaEl, BeFr, GiVa}). Note that we prefer to use the term \emph{1-form} rather than \emph{covector} in what follows. As well as this machinery we will need the notions of 1-form-valued 2- and 3-forms: we will define these from scratch, following closely  \citet{KaArToYaMaDe} and \cite{Fr}, in order to establish notation and properties, and because they may be unfamiliar to some readers, although such objects arise naturally in the discussion of continuum mechanics for the treatment of stress. While, as indicated above, our aim is to use purely geometrical constructions where possible, it is sometimes awkward to represent complicated contractions of objects using coordinate-free notation, and in some  calculations we will use indexed objects. Both approaches have benefits and the maximum utility is obtained by switching between them fluidly. 

\subsection{Momentum flux}

We recall that in a traditional treatment of fluid flow in Euclidean space \citep{Ba}, the stress on an element of surface with normal vector $\nv$ at a point $(\xv,t)$ is a vector force $\fv(\xv, t, \nv)$ per unit area. It can be established that $\fv$ depends linearly on $\nv$ and so we can write $f_i = \sigma_{ij}(\xv,t) n_j$, where the stress tensor $\sigmav$ is symmetric. Then, the divergence of the stress tensor $\partial_j \sigma_{ij}$ gives the net force per unit volume, and appears in the Navier--Stokes equation which in conservation form is
 \begin{equation}
\partial_t (\rho u_i) + \partial_j (\rho  u_i u_j) = \partial_j \sigma_{ij} .  
\end{equation}
This form highlights the role of the momentum flux $\rho u_i u_j$ as a tensor of a nature similar to that of $\sigmav$. For a compressible Newtonian fluid, the stress tensor is given by
 \begin{equation}
\sigma_{ij}  = - p\,  \delta_{ij} + \varsigma ( \partial_j u_i + \partial_i u_j )+ \lambda \divv u \, \delta_{ij} , 
\label{eqeucvisc} 
\end{equation}
where $p$ is the pressure field, and $\varsigma$ and $\lambda$ denote the dynamic and bulk viscosities.

In our more general setting for flow on an arbitrary three-dimensional manifold $\Mc$, the appropriate geometrical object to represent the stress is a \emph{1-form valued 2-form} $\tau$ which can be defined by 
\begin{equation}
\tau  = \tfrac{1}{2} \tau_{ijk}\,  dx^i \otimes dx^j \wedge dx^k. 
\label{eqTcdef} 
\end{equation}
This can be thought of as an object with two \emph{legs}; the first leg, given by the $i$ index, has the nature of a 1-form or covector, while the second leg, given by indices $j$ and $k$, has the nature of a 2-form. The interpretation of $\tau$ is as follows: if we have a surface element given by vectors $v$ and $w$ at a point in the fluid, and the fluid has velocity $u$ there, then the rate of working of the stress force by flow through that element of surface, per unit area, is given by contracting $\tau$ with $u$ on the first leg and $v\otimes w$ on the second leg:
\begin{equation}
\tau(u,v,w) = \tau_{ijk}\,  u^i v^j w^k. 
\label{eqTcdefcomponents} 
\end{equation}
Note that in a geometric setting momentum is a 1-form, and so it is natural to work with 1-form valued objects such as $\tau$; its \emph{value} (when contracted on the second leg) is not the force on the surface element itself, but the \emph{rate of working} or \emph{power}  of the force when contracted with the vector fluid velocity $u$ on its first leg. Nonetheless for brevity in the discussion below we refer to this 1-form value $\tau(\cdot, v, w)$ as the \emph{force}. Vector valued 2-forms, with components $\tau^i_{jk}$, may be defined similarly but we will not need these.

\subsection{Exterior covariant derivative}

Given that a 1-form valued 2-form $\tau$ is the appropriate description of the force on surface elements in a fluid flow, we need to obtain its divergence, in other words calculate a net force on elements of volume. This divergence is a \emph{1-form valued 3-form} given by $\cd \tau$, where $\cd$ is the \emph{exterior covariant derivative} defined by \citep{KaArToYaMaDe} 
\begin{equation}
( u , \cd \tau) =  d (u, \tau) - \nabla u \dwedge \tau . 
\label{eqcddef} 
\end{equation}
Here $u$ is any vector field, $(u,\tau)$ denotes $u$ contracted into the first leg of $\tau$, likewise $(u,\cd\tau)$ is $u$ contracted into the first leg of $\cd\tau$. In $\nabla u \dwedge \tau$ the $u$ is contracted into the first leg of $\tau$ and the covariant derivative is wedged with the second leg of $\tau$. In the general use of $\dwedge$, the first legs of the two sides are contracted, the second legs are wedged: for example  for 1-forms $\alpha$ and $\beta$, a 2-form $\gamma$ and a vector $u$,
\begin{equation}
(u  \otimes \alpha ) \dwedge (\beta\otimes\gamma) = (u,\beta) \, \alpha \wedge \gamma.
\end{equation}
Consistent with this, we adopt the (somewhat awkward) convention that the first leg of $\nabla u$ is taken to be $u$ and the second to be $\nabla$ and write
\begin{equation}
\nabla u = \nabla_j u^i  \, \partial_i \otimes dx^j= u^i_{\ ;j}   \, \partial_i \otimes dx^j , 
\end{equation}
with a semicolon as alternative notation for a covariant derivative \citep{KaArToYaMaDe}. 

Using components the definition of $\cd$ amounts to 
\begin{equation}
( u , \cd \tau)_{ijk} =  u^m  (\cd \tau)_{mijk} = 3 ( u^m \tau_{m[ij})_{;k]} - 3  \tau_{m[ij} \nabla_{k]} u^m = 3 u^m  \tau_{m[ij;k]}
\end{equation}
and so we have 
\begin{equation}
(\cd \tau) _{mijk} = 3  \tau_{m[ij;k]}, 
\label{eqcdindices}
\end{equation}
with square brackets denoting antisymmetrisation (see \cite{Sc80} for the formulation of exterior derivatives and wedge products in terms of antisymmetrised tensors). The definition is thus independent of the choice of $u$. 
The resulting object $\cd\tau$ has the physical interpretation that the net force on a volume element supplied by vectors $u$, $v$ and $w$ is the 1-form obtained as the first leg of $\cd\tau$, when we take the contraction $\cd \tau(\cdot, u,v,w)$ on the second leg. 
We note that the appearance of the covariant derivative in this definition is inevitable, since computing the net force on a volume element involves the differences between forces on the various faces and taking these differences requires parallel transport. 

The above general definition (\ref{eqcddef}) of $\cd \tau$ in fact holds for 1-form valued $p$-forms for any $p$ and is easily extended to $p$-forms with values in other vector bundles. The theory of these `valued' forms and the exterior covariant derivative $\cd$ was developed by E. Cartan as the natural language for discussing curvature, gauge theories, and  stress in elasticity and fluid flow \citep{Fr,KaArToYaMaDe}.

The usual operations such as raising and lowering indices with $\sharp$ and $\flat$, and  the Hodge star $\star$ operator can be applied to either leg of $\tau$, with a numeral subscript used to indicate which leg. With this notation, we can relate $\tau$ to the usual definition of the (twice contravariant) stress tensor $T = T^{ij}\partial_i \otimes \partial_j$ through
\begin{equation}
\tau = \star_2  \flat_2\flat_1 \  T, 
\label{eqtauT}
\end{equation}
or in components
\begin{equation}
\tau_{ijk} = g_{il} \, T^{lm} \, \mu_{mjk} . 
\end{equation}
We also need to relate the  exterior covariant derivative of $\tau$ to the usual divergence of the  tensor $T$. We have that
\begin{equation}
(\cd \tau)_{mijk} = 3 ( g_{ml}\,  T^{ln} \, \mu_{n[ij} )_{;k]} = 3 g_{ml} \, \mu_{n[ij} \, T^{ln}_{\,\,;k]} , 
\end{equation}
as the covariant derivatives of $g$ and $\mu$ vanish. 
A short computation shows this reduces to
\begin{equation}
(\cd \tau)_{mijk} =  g_{ml}  \, T^{ln}_{\,\,;n}\,  \mu_{ijk} . 
\label{eqcdcovariant} 
\end{equation}
This is precisely $\cd\tau = \alpha \otimes \mu$ with $ \alpha_m = g_{ml}  \, T^{ln}_{\,\,;n} $, giving the natural relation between the 1-form valued 3-form $\cd \tau$ and the usual  divergence $T^{ij}_{\,\,;j}$ of $T^{ij}$. 

The symmetry of the stress tensor, easily expressed as  $T^{ij}=T^{ji}$ or $T(\alpha,\beta)=T(\beta,\alpha)$ for arbitrary 1-forms $\alpha$ and $\beta$, can be rewritten in terms of the 1-form valued 1-form $\star_2\tau = T_{ij} d x^i \otimes d x^j$ as
\begin{equation} 
{\star_2 \tau} (u,v) = \star_2 \tau (v,u)
\label{eqtausym}
\end{equation}
for arbitrary vectors $u$ and $v$. It can  equivalently be  stated in terms of $\tau$ itself as
\begin{equation}
({\alpha}_\sharp  \otimes \beta) \dwedge \tau = ({\beta}_\sharp \otimes \alpha) \dwedge \tau,
\label{eqtausym2}
\end{equation}
for arbitrary 1-forms $\alpha$ and $\beta$, 
by applying the property that for any 2-form $\gamma$, 
%
\begin{equation}
\beta \wedge \gamma =  (\beta_\sharp ,\star \gamma) \, \mu, 
\label{eq:prop0}  
\end{equation}
to the second leg of $\star_2 \tau$. 
 



\subsection{Interpretation}

For a useful physical interpretation of the definition (\ref{eqcddef}) of $\cd \tau$, consider the work done by the stress $\tau$ on the surface of a volume $\Vc$ moving with a velocity field $u$. The rate of work, that is the power generated, is given by
\begin{equation}
P = \int_{\partial \Vc} (u,\tau) = \int_{\Vc} d(u,\tau) = \int_\Vc (u,\cd \tau) + \int_\Vc \nabla u \dwedge \tau,
\label{eqpower}
\end{equation}
where, as usual, the contraction in $(u,\tau)$ is into the first leg of $\tau$.
The first term on the right-hand side corresponds to the work done by the force $\cd \tau$ on the moving volume $\Vc$ and is associated with a change in kinetic energy; the second term corresponds to an internal work and is associated with the deformation of $\Vc$ and the resulting change of internal energy. This is better seen by rewriting the second term as
\begin{equation}
\int_\Vc \nabla u \dwedge \tau = \tfrac{1}{2} \int_\Vc (\sharp_1 \lie_u g)  \dwedge \tau = \tfrac{1}{2} \int_\Vc \langle\!\langle\lie_u g,\star_2 \tau\rangle\!\rangle \mu,
\label{eqinterwork}
\end{equation}
where $\lie_u$ denotes the Lie derivative along $u$ and  $\langle \! \langle \cdot, \cdot \rangle \! \rangle$ denotes the contraction of tensors defined, using the metric twice, as $\langle\!\langle\sigma,\tau\rangle\!\rangle = g^{ij} g^{kl} \sigma_{ik} \tau_{jl} $. 
We have also used the result  
\begin{equation}
\tfrac{1}{2} \lie_u g = \nabla u_\flat + \tfrac{1}{2} d u_\flat = \tfrac{1}{2} \bigl( \nabla u_\flat + (\nabla u_\flat)^{\mathrm{T}} \bigr),
\label{eqniftyidentity} 
\end{equation}
that is, $\tfrac{1}{2} \lie_u g$ is the symmetrisation of $\nabla u_\flat$. This follows from the computation
\begin{align}
\lie_u g(v,u)&=\lie_u(g(v,w))-g(\lie_u v,w)-g(v,\lie_w) = \nabla_u g(v,w) -g(\lie_u v,w)-g(v,\lie_w)  \nonumber \\
&=g(\nabla_u v - \lie_u v,w) + g(v,\nabla_u w -\lie_u w) = (\nabla_v u_\flat)(w) + (\nabla_w u_\flat)(v) \nonumber \\
&= 2 (\nabla_w u_\flat)(v) +  (\nabla_v u_\flat)(w) - (\nabla_w u_\flat)(v) = 2 (\nabla u_\flat)(v,w) + d u_\flat(v,w),
\end{align}
for arbitrary vectors $v, \, w$, using that $\nabla_u g(v,w)= g(\nabla_u v,w)+g(v,\nabla_u w)$ and $\lie_u v = \nabla_u v - \nabla_v u$.
We emphasise that $\lie_u g$ provides a natural measure of the deformation induced by $u$, consistent with the interpretation of  (\ref{eqinterwork}) as the power associated with the deformation of $\Vc$.
  
For vector fields $u$ that satisfy $\nabla u =0$, and so are parallel-transported across $\Mc$, (\ref{eqpower}) reduces to 
\begin{equation}
\int_{\Vc} (u, \cd\tau) = \int_{\Vc} d(u,\tau)  \quad \qquad (\nabla u = 0), 
\end{equation}
which gives a metric-independent weak form of $\cd \tau$ that can be exploited for momentum-preserving discretisation \citep{ToHuGe,Ge}.

\subsection{Properties of $\cd$}

We conclude this section with properties of the exterior covariant derivative $\cd$ useful for our purpose. We first note that we can regard any  3-form $\omega$ as a 1-form valued 2-form by simply using the formula (\ref{eqTcdefcomponents}), and with this in mind it is easy to establish that when multiplied by a scalar function $f$ we have 
\begin{equation}
\cd (f \omega) =  df \otimes \omega  + f \, \cd \omega. 
\label{eqcdLeib}
\end{equation}
In addition, when $\omega$ is the metric-induced volume form $\mu$ on $\Mc$ it follows from $\nabla\mu=0$ that 
\begin{equation}
\cd \mu = 0 . 
\label{eqcdmu}
\end{equation}

In writing the equations of fluid mechanics in a general setting, Lie derivatives naturally emerge that express transport of quantities. For example in the Euler equation (\ref{eqeuler}) below, a Lie derivative $\lie_u \nu$ appears to express transport of momentum, in place of the traditional $\uv\cdot \nabla \uv$ in Euclidean space. Thus crucial to any analysis is a link between the divergence $\cd$ of a quantity and an appropriate Lie derivative.
We use the following key identity, which holds for any vector field $u$, 1-form field $\alpha$ and 3-form field $\omega$,  
\begin{equation}
\lie_{u} ( \alpha \otimes \omega) = \cd (\alpha \otimes u \ip \omega) + (\nabla u , \alpha) \otimes \omega, 
\label{eqcdform1} 
\end{equation}
and links a Lie derivative of the 1-form valued 3-form $\alpha\otimes\omega$ and the exterior covariant derivative of the 1-form valued 2-form $\alpha \otimes u \ip \omega$. In the term $(\nabla u,\alpha)$ the inner product is taken between the $u$ and the $\alpha$, leaving behind a 1-form. To prove this identity we contract the left-hand side with an arbitrary vector field $v$ on the first leg only, so that for example $(v,\alpha\otimes \omega) = (v,\alpha)\omega$, writing first
\begin{align}
(v, \lie_u (\alpha\otimes \omega) )  &  = \lie_u ( v, \alpha\otimes\omega) - ( \lie_u v , \alpha\otimes \omega)
\nonumber \\
 & = d( v, \alpha \otimes u \ip \omega)  - ( \lie_u v , \alpha\otimes \omega) ,
\end{align}
using Cartan's formula 
\begin{equation}
\lie_{u} \beta = d(u\ip \beta) + u \ip d \beta , 
\label{eqcartan}
\end{equation}
and noting that $( v, \alpha\otimes\omega)$ is a 3-form and so vanishes under the action of $d$. 
We can now apply (\ref{eqcddef}) and $\lie_u v = \nabla_u v - \nabla_v u$ to write 
\begin{align}
(v, \lie_u (\alpha\otimes \omega) )  &  = (v, \cd( \alpha \otimes u \ip \omega) ) + \nabla v \dwedge\alpha \otimes  u \ip \omega 
- ( \nabla_u v , \alpha\otimes \omega) +  ( \nabla_v u, \alpha\otimes \omega) .
\label{eqmessylie}
\end{align}
Since 
\begin{equation}
\beta\wedge u \ip \omega  
= (\beta, u) \, \omega
\label{equsefulidentity} 
\end{equation}
for any 1-form $\beta$, letting $\nabla$ take the place of $\beta$, we observe that the second and third terms of (\ref{eqmessylie}) cancel and the last can be rewritten to give
\begin{align}
(v, \lie_u (\alpha\otimes \omega) )  &  = (v, \cd( \alpha \otimes u \ip \omega) ) + (v,  (\nabla u, \alpha) \otimes \omega) .
\label{eqliecdlink}
\end{align}
The vector field $v$ is arbitrary and so the result (\ref{eqcdform1}) follows. 

We finally observe that for practical computations, it may be preferable to avoid using the full coordinate expression (\ref{eqcdindices}) for $\cd \tau$. Instead, a convenient expression emerges by expanding $\tau$  as a sum 
\begin{equation}
\tau = d x^i \otimes \alpha^{(i)},
\end{equation}
where the $\alpha^{(i)}$ are 2-forms. The exterior covariant derivative is then given by
\begin{equation}
\cd \tau = \nabla d x^i \otimes_\wedge \alpha^{(i)} + d x^i \otimes d \alpha^{(i)},
\label{frankelstyle}
\end{equation}
where $\otimes_\wedge$ denotes a Cartesian product with the first leg of $\nabla d x^i$ and a wedge product with the second leg, with as above, the covariant derivative treated as the second leg (that is, $\nabla d x^i = \nabla_j(d x^i) \otimes d x^j$).
We can check this from the coordinate-free definition of $\cd$:
\begin{subequations}
\begin{align}
(u,\cd \tau) &= d (u^i \, \alpha^{(i)}) - \nabla u \dwedge (d x^i \otimes \alpha^{(i)}) \\
&= {u^i_{\,,j}} \, d x^j \wedge \alpha^{(i)} + u^i \,d \alpha^{(i)} - (\nabla_j u , d x^i)  \,d x^j \wedge \alpha^{(i)} \\
&=  {u^i_{\,,j}} \, d x^j \wedge \alpha^{(i)} + u^i \,d \alpha^{(i)} - ( u^i_{,j} - (u , \nabla_j \,d x^i)) \, d x^j \wedge \alpha^{(i)} \\
&= (u , d x^i \otimes d \alpha^{(i)} + \nabla_j(d x^i) \otimes  d x^j \wedge \alpha^{(i)}).
\end{align}
\end{subequations}
We illustrate the application of this formula and, more broadly, manipulations of the 1-form valued $\tau$ with explicit computations in spherical geometry in appendix \ref{app:sphere}.

\section{Application to compressible perfect fluid} 
\label{secidealflow}

Having set up the necessary machinery and linked the divergence $\cd$ to Lie derivatives, we now use this to write systems of fluid equations on a general manifold $\Mc$ in conservation form. The most fundamental case is the compressible Euler equation, which takes the coordinate-free form
\begin{subequations} \label{eulercomp}
\begin{align}
\rho [\partial_t  \nu + \lie_u \nu -  \tfrac{1}{2} \d (u,\nu) ]+ \d p  = 0 , 
\label{eqeuler}\\
\partial_t (\rho\mu) + \lie_u (\rho \mu)=0,
\label{eqcont}
\end{align}
\end{subequations}
where $\rho$ is the density,  $u$ is the velocity (vector) field, $\rho \nu  = \rho u_{\flat}$ is the corresponding (1-form) momentum and $p$ is the pressure field \citep{GiVa}. For the maximum flexibility to write a variety of fluid systems in conservation form, we develop this for the Euler equation using two distinct lines of argument.

In the first, we simply apply identities obtained in \S\ref{secmachinery} to (\ref{eqeuler}). From (\ref{eulercomp}) we can form an equation for the momentum, now thought of as the 1-form valued 3-form $\rho \nu \otimes \mu$, 
\begin{equation}
(\partial_t + \lie_u) (\rho \nu \otimes \mu) -  \tfrac{1}{2}  \rho \,  \d (u,\nu) \otimes  \mu +  \d p \otimes \mu  = 0 . 
\label{eqeuler1}
\end{equation}
We then apply (\ref{eqcdform1}) together with 
\begin{equation}
(\nabla u,\nu)=\tfrac{1}{2} \nabla (u,\nu) = \tfrac{1}{2} \d (u,\nu),
\end{equation}
as $\nu = u_\flat$ and the covariant derivative of the metric vanishes, $\nabla g = 0$, to obtain
\begin{equation}
\partial_t (\rho \nu \otimes \mu) + \cd (\rho \nu \otimes u \ip \mu ) + \d p \otimes \mu = 0. 
\end{equation}
We can also use (\ref{eqcdLeib})--(\ref{eqcdmu}), Cartan's formula and note that $u\ip \mu = \star \nu$ to write both the 
momentum and continuity equations in the desired conservation form 
\begin{subequations} \label{eqcompeuler}
\begin{align}
\partial_t (\rho \nu \otimes \mu) + \cd (\rho \nu \otimes \star \nu  +  p \mu ) &= 0,
\label{eqcompeulerconsmom} \\
\partial_t (\rho  \mu) + d ( \rho \, {\star} \nu ) &= 0.
\label{eqcompeulerconsmass} 
\end{align}
\end{subequations}
This identifies the momentum flux as the 1-form-valued 2-form $\rho \nu \otimes \star \nu$ and the mass flux as the 2-form $ \rho \, {\star} \nu$.

The second line of argument starts from an action principle \citep{GoMa,HaEl} and provides a direct variational derivation of the Euler equations in conservation form, as an alternative to the Euler--Poincar\'e derivation which yields (\ref{eqeuler}) \citep{Ne62,Sa88,HoMaRa,We18,GiVa} and which we record in Appendix \ref{app:EP} for completeness.
We suppose that the time-dependent family of diffeomorphisms $\phi_t: \Mc \to \Mc$ moves the fluid elements, together with the mass  3-form $\rho\mu$ and the scalar entropy $s$, from some initial configuration. If we let the internal energy be $e(\rho, s)$ per unit mass, the action is given by 
\begin{equation}
\Ac[\phi] =  \int dt \int_{\Mc}    L [\phi], \quad \textrm{where} \quad L[\phi]=  \left[ \tfrac{1}{2} g(u, u) -  e(\rho,s) \right]   \rho\mu
\label{eqaction} 
\end{equation}
is the Lagrangian 3-form, that is the Lagrangian density multiplied by $\mu$. Here
we abbreviate $\phi$ for $\phi_t$ and 
\begin{equation}
u  = \dot{\phi} \circ \phi^{-1}, \quad \rho \mu = \phi_*  (\rho_0 \mu), \quad s = \phi_* s_0, 
\end{equation}
where $\phi_*$ is the push forward under the map $\phi$ from the initial conditions, with $\rho_0$ as the initial density, $s_0$ the initial entropy. 

We require the action to be stationary under any variation $\phi \mapsto \psi_\eps \circ \phi$, where $\psi_\eps$ is a family of mappings with $\psi_0$ the identity, so that
\begin{equation}
\dt{}{\eps} \Big|_{\eps=0}  \, \Ac [ \psi_\eps \circ \phi ] = 0 . 
\label{eqactionvary}
\end{equation}
We can take the family $\psi_\eps$ to be generated by a vector field $w$ at $\eps=0$. We can choose $w$ to vanish except between some initial and final time, and to vanish outside some local region of $\Mc$, meaning that we can freely integrate by parts in time or on $\Mc$ in what follows. Under such a variation we obtain variations in the fields, labelled fleetingly  by $\eps$, with
\begin{subequations}
\begin{align} 
& \left. \dt{}{\eps} \right|_{\eps=0} u_\eps = \partial_t w + \lie_u w = \partial_t w - \lie_w u , \label{eqvaryu} \\
& \left. \dt{}{\eps} \right|_{\eps=0} \rho_\eps\mu  = - \lie_w  (\rho \mu)  = - \divv (\rho w) \, \mu, \\
& \left. \dt{}{\eps} \right|_{\eps=0} \rho_\eps = - \divv (\rho w) , \\
& \left. \dt{}{\eps} \right|_{\eps=0} s_\eps = - \lie_w s  = - (ds, w) . 
\end{align}
\end{subequations}
Requiring the action (\ref{eqaction}) to be stationary, (\ref{eqactionvary}), then gives
\begin{equation}
\int dt \int_{\Mc}  \left [g(u, \partial_t w - \lie_w u) \, \rho \mu - \tfrac{1}{2} g(u,u) \,\lie_w (\rho \mu)  {+} (\rho e)_\rho \, \lie_w (\rho \mu) {+} \rho e_s  (\lie_w s ) \,\mu  \right]  = 0 , 
\label{eqvary}
\end{equation}
with the $\rho$ and $s$ subscripts denoting partial derivatives. 

The standard derivation in Appendix \ref{app:EP}  uses integration by parts to write each term in (\ref{eqvary}) as a pairing with the undifferentiated $w$ before invoking the arbitrariness of $w$ to obtain  
the Euler equations in the form (\ref{eulercomp}). To obtain the conservation form instead, we return to the action integral (\ref{eqaction}) and note that $\psi_\eps: \Mc \to \Mc$, so that we can write schematically
\begin{equation}
\Ac[\phi] 
= \int dt \int_{\psi_{\eps}  \Mc}    L [\phi] 
= \int dt \int_{ \Mc}   \psi^*_{\eps} L [\phi],
\end{equation}
with $\psi^*_\eps L$ the pull back of the Lagrangian 3-form. Differentiating with respect to $\eps$ at $\eps=0$ replaces the pull back by a Lie derivative with respect to the vector field $w$ and gives
%
\begin{equation}
 \int dt \int_{ \Mc}   \lie_w  L  [\phi] = 0 .
\label{eqlieaction}
\end{equation}
This key equation expresses the principle of covariance -- the invariance of laws of motion under change of variables -- at an infinitesimal level; it allows us to reformulate the result of applying the action principle and to obtain an equivalent form for the resulting equation of motion \citep{HaEl}.
Applying (\ref{eqlieaction})  to the  action integral (\ref{eqaction}) gives
\begin{align}
 \int dt \int_{\Mc}  \lie_w [ \tfrac{1}{2} g(u, u)\, \rho \mu  - \rho e(\rho,s)  \,\mu  ] & = \int dt \int_{\Mc}   
\big[ 
\tfrac{1}{2} (\lie_w g)(u, u)  \,\rho \mu 
+ g (u, \lie_w u) \, \rho \mu  \label{eqpushforward} \\
&  + \tfrac{1}{2} g(u, u) \, \lie_w (\rho \mu) 
- [ (\rho \,  e)_\rho  \, \lie_w \rho  + \rho e_s \, \lie_w s ] \mu {-} \rho e  \, \lie_w \mu 
\big]  = 0 . \nonumber
\end{align}
Both this equation and (\ref{eqvary}) must hold; adding them together leaves
\begin{equation}
\int dt \int_{\Mc}   
\left[ 
\tfrac{1}{2} (\lie_w g)(u, u) \, \rho \mu 
+ g (u,  \partial_t w) \, \rho \mu + p  \, \lie_w \mu \right]= 0 , 
\label{eqexp1}
\end{equation}
after simplifying and using $p = \rho^2 e_\rho$. This equation gives the momentum equation in a weak form, suitable for finite element discretisation; see \citet{ToHuGe} and \citet{Ge}.

%

We can now use integration by parts, and so discard total time derivatives or total space derivatives $d\omega$, where $\omega$ is any 2-form, by applying
\begin{equation}
\int_{\Mc} d\omega = \int_{\partial \Mc} \omega = 0 , 
\label{eqstokestheorem} 
\end{equation}
given that $\omega$ vanishes on the boundary $\partial\Mc$. This typically requires boundary conditions on the fields, here that $u$ be parallel to $\partial \Mc$, and using that $w$, as the flow generating a diffeomorphism from $\Mc$ to $\Mc$ is also parallel to $\partial \Mc$. We denote the equivalence up to total time and space derivatives by $\simeq$. For the last two terms in (\ref{eqexp1}) we find
\begin{subequations}
\begin{align}
g(u, \partial_t w)\, \rho \mu  &= ( \partial_t w  , \nu \otimes \rho \mu ) \byparts - (w , \partial_t ( \rho\nu \otimes  \mu) ), \label{uw_t} \\
p  \, \lie_w \mu & \byparts  { - \mu \, \lie_w  p = } - (w,dp) \, \mu = - (w,dp \otimes \mu),
\end{align}
\end{subequations}
on using that $\lie_w p = (w,dp)$.
For the first term we claim that 
\begin{equation}
\tfrac{1}{2} (\lie_w g)(u, u)\, \rho \mu \byparts - ( w, \cd  (\rho\nu \otimes \star\nu)) .
\label{eqliecdidentity} 
\end{equation}
Substituting into (\ref{eqexp1}) then gives 
\begin{equation}
\int dt \int_{\Mc}   
\left[ 
( w, \cd  (\rho\nu \otimes \star\nu))  
+ (w, \partial_t (\rho \nu \otimes  \mu)) + (w , dp \otimes \mu) \right]= 0 , 
\label{eqexp1a}
\end{equation}
and as the vector field $w$ is arbitrary (albeit parallel to $\partial \Mc$), the conservation form (\ref{eqcompeulerconsmom}) must hold, completing the derivation directly from the action principle. We remark that the covariance of the action (\ref{eqlieaction}) merely encodes an identity, namely (\ref{eqcdform1}), in the form used to go from the advective to the conservation forms of the momentum equation. Its benefit lies in the cancellations of terms that arise when it is added to the stationarity condition of the action, that is, when (\ref{eqvary}) and (\ref{eqpushforward}) are added together. These cancellations are a generic feature of the approach, as the consideration of an abstract model in \S\ref{sec:abstract} demonstrates.

We now need to prove the identity (\ref{eqliecdidentity}). First we use the identity (\ref{eqniftyidentity})
contracted with the symmetric tensor $u \otimes u $ to write
\begin{equation}
\tfrac{1}{2} (\lie_w g)(u, u)  =( \nabla w_{\flat}) (u, u)  = (u ,  \nabla_u  w_{\flat} )   = (\nu , \nabla_u w)  , 
\end{equation}
using that $\nabla g = 0$. Then we have, applying (\ref{equsefulidentity}) to the contraction between the $u$ and the $\nabla$,  
\begin{equation}
\tfrac{1}{2} (\lie_w g)(u, u)  \, \rho \mu = (\rho \nu , \nabla_u w)    \, \mu  = \nabla w \dwedge \rho\nu \otimes  u \ip \mu  = \nabla w \dwedge \rho\nu \otimes  \star\nu . 
\end{equation}
Hence by the definition of $\cd$, and discarding the resulting divergence term (as per integration by parts), we have 
\begin{equation}
\tfrac{1}{2} (\lie_w g)(u, u) \,  \rho \mu = 
- ( w, \cd  (\rho\nu \otimes \star\nu))  +  d(w, \rho\nu \otimes \star\nu) \byparts - ( w, \cd  (\rho\nu \otimes \star\nu)), 
\end{equation}
which establishes (\ref{eqliecdidentity}).

\section{Other fluid models}  \label{secothers}

The above calculation establishes the principle that allows us to obtain equations in conservation form by playing off the terms gained from the variational principle in (\ref{eqactionvary}) with those obtained by an infinitesimal change of variables in the integral, the limiting Lie derivative action of a pull back, in the covariance condition (\ref{eqlieaction}). This systematic method can be applied to other systems, with varying level of complexity in the resulting calculations. We consider three important specific systems, namely incompressible fluid flow, the Euler-$\alpha$ model and MHD before illuminating the overall structure by examining an abstract model of Euler--Poincar\'e type.  

\subsection{Incompressible perfect fluid}  \label{secincomp}

We commence with the Euler equations for an incompressible fluid. The action in this case takes the form
\begin{equation}
\Ac[\phi,\pi ] 
= \int dt \int_{\Mc}  \left[  \tfrac{1}{2} g(u, u)  \mu  - \pi  ( \phi_* \mu -\mu) \right], 
\label{eqeuleraction}
\end{equation}
where  $-\pi$ is a Lagrangian multiplier enforcing the volume-preservation constraint  $\phi_* \mu = \mu$.
Under variation of the path, we obtain
\begin{equation}
\dt{}{\eps} \Big|_{\eps=0}  \, \Ac [ \psi_\eps \circ \phi,\pi ] = \int dt \int_{\Mc} \left[ g(u, \partial_t w - \lie_w u)  \mu + \pi \lie_w  (\phi_* \mu-\mu) + \pi  \lie_w \mu   \right] = 0, 
\label{eqvaryincomp1}
\end{equation}
while the covariance condition (\ref{eqlieaction}) gives
\begin{equation}
\int dt \int_{M}   
\left[ 
\tfrac{1}{2} (\lie_w g)(u, u)\, \mu 
+ g (u, \lie_w u)\,  \mu
+ \tfrac{1}{2} g(u, u)  \, \lie_w \mu - ( \lie_w \pi) (\phi_* \mu-\mu) \right] = 0 , 
\label{eqpushforwardincomp1}
\end{equation}
which holds for any map $\phi$ and field $\pi$. We now impose the incompressibility condition $\phi_* \mu = \mu$ (as follows from variations of (\ref{eqvaryincomp1}) in $\pi$) in the integrals above which become 
\begin{align}
& \int dt \int_{\Mc} \left[ g(u, \partial_t w - \lie_w u) \, \mu + \pi  \, \lie_w \mu  \right] = 0 , 
\label{eqEulervary} \\
&  \int dt \int_{M}   
\left[ 
\tfrac{1}{2} (\lie_w g)(u, u)  \, \mu 
+ g (u, \lie_w u) \, \mu
+ \tfrac{1}{2} g(u, u) \,  \lie_w \mu 
\right] = 0 . 
\label{eqpushforwardincomp2}
\end{align}
As before we add these two equations to obtain
\begin{equation}
\int dt \int_{\Mc} \left[ 
\tfrac{1}{2} (\lie_w g)(u, u) \, \mu  + 
g(u, \partial_t w )\,  \mu + (\pi +  \tfrac{1}{2} g(u, u) )\, \lie_w \mu  \right] = 0 . 
\label{eqniceeuler}
\end{equation}
If we set $p = \pi +   \tfrac{1}{2} g(u, u)$,  we recover (\ref{eqexp1}) with $\rho=1$ and, following the compressible case, the incompressible equations in  the form
\begin{subequations} \label{eqeulera}
\begin{align}
\partial_t ( \nu \otimes \mu) + \cd (\nu \otimes \star \nu  +  p \mu ) &= 0,
\label{eqeulerconsmom} \\
\divv u  &= 0,
\label{eqeulerconsmass} 
\end{align}
\end{subequations}
with $\mu \divv u  = d \, {\star} \nu$. 

\subsection{Euler-$\alpha$ model}  \label{secalpha}


We next consider the Lagrangian averaged Euler-$\alpha$ model first introduced by \citet{Ho99}. The model is a generalisation of the Euler equations for incompressible perfect fluids that accounts for the averaged effect of small-scale fluctuations (see \citet{Ho02a}, \citet{MaSh}, \citet{Ol} and \citet{OlVa19} for increasingly sophisticated heuristic derivations); it has  been formulated on Riemannian manifolds \citep{MaRaSh,Sh98,Sh00,GaRa,OlVa19}. We now show that the variational route enables a relatively straightforward derivation of the conservation form of the Euler-$\alpha$ model on manifolds, which otherwise would be difficult to obtain.

The Euler-$\alpha$ action for an incompressible flow $u$ is 
\begin{equation}
\Ac[\phi] = \int dt \int_{\Mc}  \bigl[ \tfrac{1}{2} g(u, u)\,  \mu  + \tfrac{1}{4} \alpha^2 |\lie_u g|^2\,  \mu  - \pi  ( \phi_* \mu -\mu) \bigr]  
\label{eqactionalpha} 
\end{equation}
where $\alpha$ is a parameter and $|\lie_u g|^2 = \langle\!\langle \lie_u g,\lie_u g\rangle\!\rangle$ is the square of the deformation of $u$ (cf.\ (\ref{eqniftyidentity})).  This action is identical to Euler action (\ref{eqeuleraction}) except for the addition of the middle term, which we denote by $\alpha^2 \Ac_2$. We note that other forms for this term -- equivalent in Euclidean geometry but distinct on curved manifolds -- have been proposed originally  \citep{MaRaSh,Sh98} and that (\ref{eqactionalpha}) follows the more recent literature \citep{Sh00,GaRa,OlVa19}.
We focus on $\alpha^2 \Ac_2$ since we have dealt with the other two terms in the treatment of the Euler equations above. For simplicity, we assume that the manifold $\Mc$ has empty boundary to avoid unnecessary complications when discarding integrals over $\Mc$ that are the derivative $d$ of a 2-form (see \cite{Sh00} for a careful treatment of the boundary conditions). We have
\begin{subequations}
\begin{align}
\Ac_2[\phi] &= \tfrac{1}{4} \int dt \int_{\Mc} \langle\!\langle \lie_u g,\lie_u g\rangle\!\rangle\, \mu  = \tfrac{1}{2}  \int dt \int_{\Mc} \langle\!\langle\nabla u_{\flat} ,\lie_u g\rangle\!\rangle \,\mu  \label{eqactionalpha1}\\
&= \tfrac{1}{2}  \int dt \int_{\Mc} \nabla u \dwedge \star_2 \lie_u g  =  - \tfrac{1}{2}  \int dt \int_{\Mc} (u, \cd (\star_2 \lie_u g))  
= - \tfrac{1}{2}  \int dt \int_{\Mc} (u, \Delta_R \nu) \, \mu \label{eqactionalpha2}
\end{align}
\end{subequations}
on using (\ref{eqcddef}), (\ref{eqniftyidentity})  and   (\ref{eqliecdlink}). In the last equality, we have introduced the \emph{Ricci Laplacian} of  1-forms via
\begin{equation}
\Delta_R \nu \otimes \mu = \cd(  \star_2 \, \lie_u g),
\label{eqricci}
\end{equation} 
recalling that $\nu = u_\flat$.
This is related to the Laplace--de Rham operator $\Delta \nu = - ( {\star} d{\star} d + d{\star} d{\star} )\nu$ and the analyst's (or rough) Laplacian 
$(\tilde\Delta \nu )_i = g^{jk} \nabla_j \nabla_k \nu_i$ through
\begin{equation}
 \Delta_R \nu = \Delta \nu + 2R(u)   = \tilde\Delta \nu + R(u), 
 \label{eqmanylaplacians}
\end{equation}
where $R$ is the Ricci tensor given by, in general, $R(u)_i = R_{ij} u^j = \nabla_j \nabla_i u^j  - \nabla_i \nabla_j u^j$. 
The latter equality in (\ref{eqmanylaplacians}) is kown as the Weizenb\"ock formula \citep{Fr}; we check the former. 
%
%
Setting temporarily $S_{ij} = (\lie_u g)_{ij}$, (\ref{eqcdcovariant}) shows that we need to compute $\nabla_j S^{ij}$, which gives
\begin{subequations}
\begin{align}
\nabla_j S^{ij} & = g^{ik}\,  g^{jl} \, \nabla_j (\lie_u g)_{kl} = g^{ik} \, g^{jl} \, \nabla_j (\nabla_k u_l + \nabla_l u_k )  \\
& = g^{ik} \, [ \nabla_j \nabla_k u^j + ( \tilde{\Delta} \nu)_k  ]= g^{ik} \, (R(u) + \tilde{\Delta} \nu)_k,  
\end{align}
\end{subequations}
using incompressibility, $\divv u = \nabla_i u^i = 0$.

%

The Euler--$\alpha$ momentum equation is obtained by extremising the action (\ref{eqactionalpha}) under variations of the form (\ref{eqvaryu}). The contribution of $\Ac_2$ is readily obtained from (\ref{eqactionalpha2})  using the self-adjointness of $\Delta_R$ (as used in \cite{OlVa19}) to find
\begin{equation}
\left. \dt{}{\eps} \right|_{\eps=0} \Ac_2 = - \int dt \int_{\Mc}  (\partial_t w - \lie_w u, \Delta_R  \nu ) \, \mu .
\label{eqalphavary} 
\end{equation}
Adding this to the variation obtained for the Euler equation in (\ref{eqEulervary}) and requiring the sum to vanish for arbitrary $w$ yields the Euler--$\alpha$ equations in the advective form
\begin{equation}
\partial_t \upsilon + \lie_u \upsilon + d \pi = 0 , \quad \divv u = 0 , \quad \textrm{where} \quad \upsilon = \nu - \alpha^2 \Delta_R \nu  . 
\label{eqeuleralphaeq} 
\end{equation}


It is not obvious how to put (\ref{eqeuleralphaeq}) into conservation form by inspection and so we proceed to use the pull back of the action according to (\ref{eqlieaction}). We focus again on $\Ac_2$ since the contributions of the other terms are as in (\ref{eqpushforwardincomp2}). The variation of $\Ac_2$ can be written as the sum of three terms  proportional to $\lie_w u$, $\lie_w g$ and $\lie_w \mu$. It is convenient to use the form (\ref{eqactionalpha2}) of $\Ac_2$ for the first and (\ref{eqactionalpha1}) for the other two. This leads to
\begin{equation}
\int dt \int_{\Mc} \lie_w L_2[\phi] = \int dt \int_{\Mc}  \left[ (- \lie_w u, \Delta_R  \nu )\,  \mu + \tfrac{1}{4} \tilde{\lie}_w  |\lie_u g|^2 \, \mu + \tfrac{1}{4}   |\lie_u g|^2\,  \lie_w \mu \right],
\label{eqpushforwardalpha}
\end{equation}
where the tilde in $\tilde{\lie}_w$ indicates  a Lie derivative at fixed $u$. We work out the second term in coordinates, noting that, as $g_{ij} \, g^{jk} = \delta_i^k$,
\begin{equation}
 \lie_u ( g^{ij} ) = - g^{ik} \, g^{lj} \, (  \lie_u g_{kl} )  =  - g^{ik} \, g^{lj}  \,  ( \lie_u g)_{kl}  
 \equiv - (\lie_u g)^{ij} , 
 \label{eginv}
\end{equation} 
to obtain
\begin{subequations}
\begin{align}
 \tilde{\lie}_w  |\lie_u g|^2  &=  \tilde{\lie}_w \big[ g^{ik} \, g^{jl} \,  (\lie_{u} g)_{ij} \, (\lie_u g)_{kl} \big] \nonumber \\ 
 &= -2 (\lie_w g)^{ik}\,  g^{jl} \,  (\lie_{u} g)_{ij} \, (\lie_u g)_{kl} + 2 g^{ik} \, g^{jl}  \, (\lie_{u} \lie_w g)_{ij}\,  (\lie_u g)_{kl}  \nonumber \\
&\simeq -2 (\lie_w g)_{ik}\,  (\lie_{u} g)^{ij} \, (\lie_u g)_{j}^{k} + 4 (\lie_w g)_{ij} \, (\lie_u g)^{ik}  \,  (\lie_u g)^j_k 
- 2 (\lie_w g)_{ij} \, g^{ik} \, g^{jl} \,  (\lie_{u}  \lie_u g)_{kl} \nonumber \\
& = 2 \, \langle\!\langle \lie_w g,T\rangle\!\rangle,
\end{align} 
\end{subequations}
where we introduce the twice covariant tensor
\begin{equation}
T =  (\lie_u g)^2 - \lie_u \lie_u g, \quad \textrm{i.e.} \quad  T_{ij} =  g^{kl}\,  (\lie_u g)_{ik} \, (\lie_u g)_{jl} - ( \lie_u\lie_u g)_{ij}. 
\label{eqnicealpha}
\end{equation}
Adding together the variations (\ref{eqvaryincomp1}), (\ref{eqpushforwardincomp1}), (\ref{eqalphavary}) and (\ref{eqpushforwardalpha}) then leads to
\begin{align}
\int dt \int_{\Mc} & \Big[ 
\tfrac{1}{2} (\lie_w g)(u, u) \, \mu  +  g(\partial_t w,u )\,  \mu + p \, \lie_w \mu  
\nonumber \\
 & \left.  - \alpha^2 (\partial_t w,\Delta_R \nu) \, {\mu} + \tfrac{1}{2} \alpha^2  \langle\!\langle   \lie_w g,T\rangle\!\rangle   {\mu} +  \tfrac{1}{4}  \alpha^2  |\lie_u g|^2 \,  \lie_w \mu \right] = 0 .
 \label{eqalphabyparts}
\end{align}
Integrating by parts, in particular using that
\begin{equation}
\tfrac{1}{2} \langle\!\langle\lie_w g,T\rangle\!\rangle {\mu} =\nabla w \dwedge \star_2 T \byparts  {-}  (w,\cd \,{\star}_2 T),
\end{equation}
and requiring (\ref{eqalphabyparts}) to vanish for arbitrary $w$ gives  the conservation form of the Euler--$\alpha$ equation,
\begin{equation}
\partial_t (\upsilon \otimes \mu) + \cd \left[ \nu \otimes {\star} \nu + \alpha^2 \bigl( {\star}_2 T +  \tfrac{1}{4} |\lie_g u|^2  \mu \bigr) + p  \mu \right]   = 0 .
\label{eqeulerfinal}
\end{equation}
A direct check that this can be expanded to give (\ref{eqeuleralphaeq}) is tedious but confirms the result. We emphasise that the momentum flux tensor that emerges as the argument of $\cd$ is not simply $\upsilon \otimes {\star} \nu = \upsilon \otimes u \ip \mu$, namely transport of the momentum $\upsilon$ by the velocity $u$,  as might have been expected naively. The latter tensor is not symmetric, whereas the tensor we obtain in (\ref{eqeulerfinal}) is symmetric by construction \citep{HaEl,GoMa}. Note that the pressure is augmented by the fluctuations giving the total effective pressure as $p + \tfrac{1}{4} \alpha^2 |\lie_g u|^2 $.

\subsection{Magnetohydrodynamics} \label{secmhd}

Finally we consider magnetohydrodynamics (MHD) and outline a derivation of the conservation form of the governing equation of ideal MHD which generalises (\ref{eqcompeuler}) by including the Lorentz force; see the classic study by \cite{Ne62} and also \cite{GiVa19b}. The general procedure is already established, but because the flow $u$ and and magnetic field $b$ have distinct transport properties, there are notable differences, and one effect is that a magnetic pressure term emerges from the analysis.

The MHD action is given by $\Ac - \Bc$ where $\Ac$ is the compressible perfect fluid action (\ref{eqaction}) and
\begin{equation}  
\Bc[\phi] =    \int dt \int_{\Mc}   \tfrac{1}{2} g(b , b ) \, \mu  
\label{eqBc}
\end{equation}
is the magnetic energy. Here $b$ is the magnetic vector field,  and we again allow $\Mc$ to have a non-empty boundary with the boundary condition $b \parallel \partial \Mc$. The most fundamental representation of the magnetic field is perhaps not the vector field $b$ itself but the associated magnetic flux 2-form, $\beta = b \ip \mu$ \cite{Fr}. The absence of magnetic monopoles, that the flux across any closed surface is zero, is simply expressed by $\beta$ being closed, $d\beta=0$ and hence $\divv b = 0$. The flux 2-form is transported by the flow so that
\begin{equation}
\partial_t \beta + \lie_u \beta =0, 
\end{equation}
or equivalently  pushed forward from the initial condition according to $\beta = \phi_* \beta_0$. The magnetic vector field $b$ obeys a more complicated equation \citep[and in fact may be considered as a tensor density; see][]{RoSo06b},
\begin{equation}
\partial_t b + \lie_u  b  + b \divv u =0.
\label{eqbtensorweight}  
\end{equation}

Let us now consider the effect of a variation in the path $\phi \mapsto\psi_\eps \circ \phi$ on $\mathcal{B}$ {\citep{Ne62}}. We have using (\ref{eqbtensorweight}) that $b$ is transported according to 
\begin{align} 
\left. \dt{}{\eps} \right|_{\eps=0} b_\eps &=  - \lie_w b - (\divv w) \, b    , 
\label{eqtransb} 
\end{align}
and so making the total action $\Ac-\Bc$ stationary introduces new integral terms:
\begin{equation}
 \left. \dt{}{\eps} \right|_{\eps=0}\Bc [ \psi_\eps\circ\phi] 
 = \int dt \int_{\Mc}  [ - g (b , \lie_w b)\,  \mu  - g( b,b)\,  \lie_w \mu  ] . 
 \label{eqBactionvary}
\end{equation}
Combining with $\mathrm{d} \Ac / \mathrm{d} \eps |_{\eps=0}$ in  (\ref{eqvary}), and using the integration by parts identities (\ref{eqbyparts}) and similar, gives the momentum equation
\begin{equation}
\partial_t (\nu\otimes \rho \mu) + \lie_u ( \nu \otimes \rho \mu) - \tfrac{1}{2} d (\nu,u) \otimes \rho \mu +  dp \otimes \mu  =  \lie_b (\star \beta\otimes \mu) -  dg(b,b) \otimes \mu,
\label{eqmhd}
\end{equation}
noting that $b_\flat = \star \beta$.

To obtain the conservation form of (\ref{eqmhd}), we use the covariance of the action (\ref{eqlieaction}), adding to (\ref{eqBactionvary}) the term
\begin{equation}
\int dt \int_{\Mc}  \left[   
 \tfrac{1}{2} (\lie_w g) (b , b )\, \mu  + 
  g (b , \lie_w b )\, \mu   +
 \tfrac{1}{2} g (b , b )\, 
 \lie_w\mu 
 \right] = 0 .
  \label{eqbpush}
\end{equation}
This gives 
\begin{equation}
  \left. \dt{}{\eps} \right|_{\eps=0}\Bc [ \psi_\eps\circ\phi] =   \int dt \int_{\Mc}   \left[
  \tfrac{1}{2} (\lie_w g) (b , b)\,   \mu 
  - \tfrac{1}{2} g(b,b) \, \lie_w \mu \right].
  \label{eqniceMHD}
   \end{equation}
Subtracting this from (\ref{eqexp1}) and following the now usual manipulations we obtain the conservation form
\begin{equation}
\partial_t (\rho \nu\otimes \mu )  + \cd  (\rho\nu \otimes \star \nu + p \mu ) =  \cd (\star \beta \otimes  \beta - \tfrac{1}{2} g(b,b) \, \mu ).
\label{eqmhdcons} 
\end{equation}
The magnetic pressure term $\tfrac{1}{2} g(b,b)$ emerges naturally in the derivation, and its origin may traced back to the term $b \divv u$ in the transport equation (\ref{eqbtensorweight}) for $b$. In a compressible fluid, whereas the fundamental magnetic flux $\beta$ is simply Lie transported in the flow map, and so conserved, the magnetic vector field $b$ with $b\ip \mu = \beta$ is intensified where the fluid is locally compressed, and this contributes to increased energy density $\tfrac{1}{2} g(b,b)$ in (\ref{eqBc}) and a resulting restoring force in (\ref{eqmhdcons}). In an incompressible fluid, the magnetic pressure can simply be absorbed in the pressure $p$.
In appendix \ref{apshallow}, we also derive the shallow-water and MHD shallow-water equations in conservation form. 

\subsection{Abstract model} \label{sec:abstract}

The variational derivations above and in appendix  \ref{apshallow} indicate that combining the stationarity of the action with its covariance leads to a number of cancellations and, as a result, relatively simple expressions for the conservation and weak forms of the governing equations. To understand how these cancellations come about and illuminate the underlying structure, it is useful to consider a general, abstract fluid model of the Euler--Poincar\'e type examined by \citet{HoMaRa} and governed by the action 
\begin{equation}
\Ac[\phi] =  \int dt \int_{\Mc}    L [u,g,a], 
\end{equation}
where the Lagrangian 3-form depends on the velocity field $u$ and metric $g$, and on tensorial fields $a$ that are advected by the flow, that is, satisfy $a = \phi_* a_0$, with $a_0$ the initial fields. The stationarity of the action reads
\begin{equation}
\left. \dt{}{\eps} \right|_{\eps=0} \, \Ac[\phi_\eps]  = \int dt \int_{\Mc}  \left( 
\left(\frac{\delta L}{\delta u}\, ,   \partial_t w - \lie_w u \right)
  - \left( 
\frac{\delta L}{\delta a}\, , \lie_w a \right) \right)   = 0 
\end{equation}
using (\ref{eqvaryu}) and that $\left. \mathrm{d}  a_\eps/\mathrm{d}\eps \right|_{\eps=0} = - \lie_w \,a$. Its covariance reads
\begin{equation}
\int dt \int_{\Mc}  \lie_w \,L [u, g, a]  =  \int dt \int_{\Mc}  \left( 
\left( \frac{\delta L}{\delta u}\, , \lie_w u \right)  + \left(\frac{\delta L}{\delta g}\, , \lie_w g \right) +
\left(\frac{\delta L}{\delta a} \, , \lie_w a \right) 
\right)   = 0.
\end{equation}
Note that $\delta L/\delta g$ should be interpreted as a 3-form whose value (on a triple of vectors) is a twice contravariant tensor.
Adding the conditions yields the compact expression
\begin{equation}
\int dt \int_{\Mc}  \left( 
\left(\frac{\delta L}{\delta u}\, ,   \partial_t w \right) + \left(\frac{\delta L}{\delta g}\, , \lie_w g \right) \right)=0.
\end{equation}
We can now integrate by parts and exploit the arbitrariness of $w$. Defining the bilinear diamond operator $\diamond$ by
\begin{equation}
\int_{\Mc}  \left(S, \lie_w g \right) = - \int_{\Mc}  \left(S \diamond g, w \right) 
\end{equation}
for any tensor-valued 3-form $S$ \citep{HoMaRa,Ho02a}, we obtain the governing equation in the form
\begin{equation}
\partial_t\,  \frac{\delta L}{\delta u} + \frac{\delta L}{\delta g} \diamond g = 0.
\end{equation}

It turns out that the diamond operator $\diamond$, when applied to a pair of symmetric tensor-valued 3-form and tensor as is the case here, is equivalent to the covariant exterior derivative $\cd$. To see this, define the twice contravariant tensor $M$ (dual to $g$) by
\begin{equation}
\frac{\delta L}{\delta g} = M \otimes \mu.
\label{eq:di}
\end{equation}
Using the symmetry of $M$, (\ref{eqniftyidentity}),  (\ref{eq:prop0}) and the definition (\ref{eqcddef}) of $\cd$, we have, for any vector field $w$, 
\begin{subequations}
\begin{align}
\int_{\Mc}  \left(\frac{\delta L}{\delta g}  \diamond g, w \right) &= -  \int_{\Mc}   
\left(\frac{\delta L}{\delta g}\, ,\lie_w g \right)= 
 - \int_{\Mc}   
  \left(\lie_w g, M \right) \mu 
  =  
  - \int_{\Mc}   
  \left(\nabla w_{\flat}, M \right) \mu    \\
  &= -
   \int_{\Mc}   
  \nabla w \dwedge  {\star}_2 \flat_1\flat_2 M
   = 
   \int_{\Mc}   
   \left(w,  \cd ({\star}_2 \flat_1\flat_2 M) \right). 
  \end{align}
\end{subequations}
Hence $\delta L/\delta g \diamond g = \cd ({\star}_2 \flat_1\flat_2 M)$ and the governing equation (\ref{eq:di}) can be rewritten in the  conservation form 
\begin{equation}
 \frac{\partial}{\partial t} \, \frac{\delta L}{\delta u}   + 
  \cd ( {\star}_2 \flat_1\flat_2 M )
  = 0. 
\end{equation}
While this expression is general and pleasantly compact, obtaining the explicit form of $M$ often requires intricate computations, as our treatment of specific models illustrates, because of the complex dependence of the Lagrangian $L$ on the metric $g$, including through the volume form.  


\section{Viscosity and viscoelasticity}  \label{secviscous}

\subsection{Newtonian fluids}

We now turn to the geometric representation of the viscous stress tensor given in (\ref{eqeucvisc}) for ordinary Euclidean space.  The  construction involves the Lie derivative of the metric which, according to (\ref{eqniftyidentity}), is given by
\begin{equation}
(\lie_u g )_{ij} = \nabla_i \nu_j + \nabla_j \nu_i  = \nu_{j;i} + \nu_{i;j} , 
\end{equation}
since $\nu = u_\flat$. It is then natural to replace the terms $\partial_i u_j + \partial_j u_i$ in (\ref{eqeucvisc}) by $\lie_u g$, both following the general rule of replacing ordinary derivatives by covariant derivatives, but more importantly as in our understanding of Newtonian fluids, it is the deformation of fluid elements that generates viscous stresses, and deformation corresponds precisely to non-zero transport of the metric under a flow $u$. With this, the geometric version of the stress tensor as a 1-form valued 2-form is
\begin{equation}
\sigma = - p \mu + \varsigma \, {\star_2}  \lie_u g + \lambda (\divv u) \, \mu , 
\label{eqviscousstress}
\end{equation}
and then the Navier--Stokes momentum equation in conservation form is
\begin{equation}
\partial_t (\rho \nu \otimes \mu) + \cd (\rho \nu \otimes  \star \nu + p \mu ) =  \cd \bigl[  \varsigma \star_2  \lie_u g + \lambda (\divv u) \, \mu  \bigr].
 \label{mom4v}
\end{equation}
In the incompressible case, this simplifies as
\begin{equation}
\partial_t (\nu \otimes \mu) + \cd ( \nu \otimes  \star \nu + p \mu ) = \varsigma \Delta_R \nu,
 \label{mom4vincomp}
\end{equation}
when (\ref{eqricci}) is used to substitute the Ricci Laplacian for $\cd ( \star_2 \, \lie_u g)$ in the sole remaining viscous term. We emphasise that the Ricci Laplacian is the proper choice of Laplacian, rather than the  Laplace--de Rham operator or the analyst's Laplacian, on a manifold with non-zero Ricci tensor. This choice ensures that velocity fields that leave the metric invariant, and hence do not cause any deformation, are not dissipated, for example solid body rotation on the surface of the sphere $\Mc = S^2$ \citep{GiRiTh,LiNo}.

The total energy in the system is 
$
E = \int_{\Mc} \bigl[ \tfrac{1}{2} g(u,u) \, \rho \mu + e(\rho, s)\, \rho \mu \bigr]. 
$ 
Following the development in (\ref{eqpower})--(\ref{eqinterwork}), we can write 
\begin{align}
\frac{dE}{dt}  &= \int_{\Mc} \left[ (u,\cd \sigma) - (\rho e)_\rho \,  \lie_u (\rho \mu) - \rho e_s \,  (\lie_u s)\,  \mu \right] 
 = \int_{\Mc} d(u,\sigma') - \int_{\Mc} \nabla u \dwedge \sigma'  \nonumber \\
& = - \int_{\Mc} \tfrac{1}{2} ( \sharp_1 \lie_u g) \dwedge \sigma' 
 = -  \int_{\Mc} \tfrac{1}{2}  \langle\!\langle\lie_u g,\star_2 \sigma' \rangle\!\rangle \,  \mu,
\end{align}
where $\sigma'=\sigma + p \mu$ denotes the viscous part of the stress tensor. To obtain this we observe that the momentum flux makes no contribution to $dE/dt$, and that the terms involving the internal energy $e$ cancel out the pressure term  $-(u,dp)\,  \mu$  (after integration by parts, as in (\ref{eqaa})--(\ref{eqaaa}), and following the argument below (\ref{eqvary1})). Using the form (\ref{eqviscousstress}) of the viscous stress, we obtain
\begin{equation}
\frac{dE}{dt}  = - \int_{\Mc} \bigl[ \tfrac{1}{2}  \varsigma \langle\!\langle \lie_u g , \lie_u g \rangle\!\rangle  + \lambda (\divv u)^2    \bigr] \mu , 
\end{equation}
as $ \langle\!\langle \lie_u g , \star_2 \mu\rangle\!\rangle  = 2 \divv u$. Note that  this derivation requires the additional no-slip boundary condition $u=0$ on $\partial \Mc$ so that the term $ d(u, \star_2  \lie_u g) $ in $ d(u,\sigma)$ integrates to zero. 


%

\subsection{Viscoelastic fluids}

In models of viscoelastic fluids such as polymer solutions, the stress $\sigma$ often appears as a dynamical variable, obeying a transport equation of the form $(\partial_t + \lie_u) \sigma = \cdots$, where the right-hand side captures the rheology of the fluid. The type of tensor chosen for $\sigma$ determines the meaning of $\lie_u$, leading to different physical models depending on the choice made; standard choices take $\sigma$ as a twice covariant or a twice contravariant tensor, with the corresponding Lie derivatives termed `lower-convected' or `upper-convected' derivatives (see, e.g., \citet{MaHu}). In the context of this paper, a natural alternative takes $\sigma$ to be a 1-form valued 2-form, $\sigma = \tfrac{1}{2}\sigma_{ijk} \, d x^i \otimes d x^j \wedge d x^k$.  A coordinate expression for its Lie derivative is readily computed: since $\lie_u$ and $d$ commute, we have
\begin{align}
2 \lie_ u \sigma &= \lie_u (\sigma_{ijk})  \, d x^i \otimes d x^j \wedge d x^k + \sigma_{ijk}  \, d \lie_u(x^i) \otimes d x^j \wedge d x^k \nonumber \\
&+ \sigma_{ijk}  \, d x^i \otimes d \lie(x^j) \wedge d x^k + \sigma_{ijk}  \, d x^i \otimes d x^j \wedge d \lie_u(x^k) \nonumber \\  
&=\left[ u^l\,  \sigma_{ijk,l} + \sigma_{ljk}\,  u^l_{,i} + \sigma_{ilk}\, u^l_{,j} + \sigma_{ijl}\, u^l_{,k}  \right] d x^i \otimes d x^j \wedge d x^k,
\end{align}
where the comma indicates differentiation (see \cite{Fr} for the analogous computation for a vector valued 2-form). This derivative can be rewritten in terms of the the twice contravariant tensor $T = \sharp_1 \sharp_2 {\star}_2 \sigma$ (cf.\ (\ref{eqtauT})) but differs from the upper convected derivative by terms proportional to $\lie_u g$ that result from the lack of commutativity of $\lie_u$ with the operators $\sharp$ and $\star$.

While it is tempting to postulate an evolution equation for the 1-form valued $\sigma$ of the form $(\partial_t + \lie_u) \sigma = \cdots$ with the right-hand side containing only rheological terms, physical considerations dictate the type of the tensor that is transported by the flow and hence the form of the evolution equation. We illustrate this with a brief geometric derivation of the Oldroyd-B model \cite{Ol50} and its formulation in terms of $\sigma$. The derivation considers a solution of polymers modelled as small dumbbells whose ends are connected by springs and which move under a combination of flow motion (through Stokes drag), Hookean spring force, and thermal noise \citep{BiHaArCu}. We follow closely the presentation in \citet{MoSp}. In a continuum description, the dumbbell extension is naturally represented by a vector field, $r$ say, measuring the total extension per unit volume. The balance of the three forces then reads
\begin{equation}
\zeta (\partial_t + \lie_u) r = - 2 K r + \sqrt{4 k_B \mathcal{T} \zeta} \, \dot W,
\label{eqr}
\end{equation}
where $\zeta$ is the drag coefficient, $K$ the spring constant, $k_B$ the Boltzmann constant, $\mathcal{T}$ the temperature, and $\dot W$ a (possibly spatially dependent) vector-valued  white noise with $\langle \mathrm{d} W^i \, \mathrm{d} W^j \rangle = g^{ij} \, \mathrm{d} t$. The noise in (\ref{eqr}) is the sum of two independent white noises acting on each end of the dumbbells,  each with strength $\sqrt{2 k_B \mathcal{T} \zeta}$ as determined by the fluctuation--dissipation theorem.  
The force exerted by the dumbbells on a surface element is the spring extension $K r$ multiplied by the number of dumbbells crossing the surface. A geometrically intrinsic representation of this is simply $K r \otimes r \ip \mu$. The stress is  proportional to the average  $K \langle r \otimes r \ip \mu \rangle$ over realisations of the white noise and can be written as the 1-form valued 2-form
\begin{equation}
\sigma = K \langle r_\flat \otimes r \ip \mu \rangle - K \langle r_\flat \otimes r \ip \mu \rangle_\textrm{eq},
\label{eqrr1}
\end{equation}
where the equilibrium value is subtracted to retain only  the stress induced by the flow. 

We derive an equation for $\sigma$. Using It\^o's formula and assuming incompressibilty, $\lie_u \mu = 0$, we obtain from (\ref{eqr}) that
\begin{equation}
 (\partial_t + \lie_u) \langle r \otimes r \ip \mu \rangle = - \frac{4 K}{\zeta}  \,  \langle r \otimes r \ip \mu \rangle + \frac{4 k_B \mathcal{T}}{\zeta} \, g^{-1} \ip \mu.
 \label{eqrr2}
\end{equation}
At equilibrium, the left-hand side vanishes, leading to
\begin{equation}
\langle r \otimes r \ip \mu \rangle_\textrm{eq} = \frac{k_B \mathcal{T}}{K}\,  g^{-1} \ip \mu.
\label{eqeq}
\end{equation}
We now consider the representation of the stress in (\ref{eqrr1}) as the vector-valued 2-form
\begin{equation}
\tilde \sigma = \sharp_1 \sigma = K \langle r \otimes r \ip \mu \rangle - K \langle r \otimes r \ip \mu \rangle_\textrm{eq}.
\end{equation}
Applying $(\partial_t + \lie_u)$ and using (\ref{eqrr2}) and (\ref{eqeq}) we obtain
\begin{equation}
\lambda (\partial_t + \lie_u) \tilde \sigma + \tilde \sigma = \varsigma \,  \sharp_1 {\star}_2\,  \lie_u g,
\label{eqOB}
\end{equation}
on noting that that $\lie_u g^{-1} = - g^{-1} (\lie_u g) g^{-1}$ (see (\ref{eginv})), and that contraction with $g^{-1} \ip \mu$ amounts to an application of $\star$.
Here $\lambda={\zeta}/{4K}$ and $\varsigma={k_B \mathcal{T} \zeta}/4K$ are the relevant rheological parameters. 

Eq.\ (\ref{eqOB}) is the desired evolution equation for the stress in the Oldroyd-B model on a manifold, expressed here in terms of $\tilde \sigma$. 
It takes a more familiar form  using the usual twice contravariant stress tensor $T = \sharp_2 {\star}_2 \, \tilde \sigma$, namely
\begin{equation}
\lambda (\partial_t + \lie_u) T + T = \varsigma \, \sharp_1 \sharp_2\,  \lie_u g,
\label{eqoldroyd2}
\end{equation}
using that the operator $\sharp_2 \star_2$ involves only the volume form and hence commutes with $\lie_u$ for incompressible flows. The Lie derivative in (\ref{eqoldroyd2}) can be identified as the upper-convected derivative. Finally, the 1-form valued 2-form obeys the slightly more complicated equation 
\begin{equation}
\lambda (\partial_t + \lie_u)  \sigma +  \sigma = \varsigma \,  {\star}_2 \lie_u g + \lambda \lie_u g \ip \sharp_1  \sigma,
\end{equation}
where $( \lie_u g \ip \sharp_1 \sigma)_{ijk} =  (\lie_u g)_{il} \, g^{lm} \, \sigma_{mjk}$ in coordinates.

%
%

\section{Concluding remarks}

We conclude with three remarks.
First, one of the benefits of the conservation form of the fluid equations is that it makes the derivation of  conservation laws arising from spatial symmetries straightforward. On a manifold $\Mc$, a spatial symmetry is identified with a Killing vector field, that is, a  vector field $k$
that carries the metric without deformation, 
\begin{equation}
\lie_k \, g = 0, 
\end{equation}
or $k_{i;j} + k_{j;i} = 0$. For example, if the domain $\Mc$ is $\mathbb{R}^3$ or a periodic domain (flat torus), these are translations; for a sphere $\Mc = S^2$ these are rotations. The associated conservation law is obtained by noting that
\begin{equation}
(k, \cd \tau) = d(k, \tau) - \nabla k \dwedge \tau  =  d(k,\tau),
\label{eq:ktau}
\end{equation}
where the vanishing of the term $\nabla k \dwedge \tau$ follows from the symmetry of $\tau$ as in (\ref{eqtausym}) and use of (\ref{equsefulidentity}). Contracting $k$ with the first leg of the dynamical equation for the 1-form valued momentum
\begin{equation}
\partial_t ( \rho \nu \otimes \mu) + \cd \tau = 0
\end{equation}
then leads to the conservation law 
\begin{equation}
\partial_t ( (k,\rho \nu) \otimes \mu) + d (k,\tau) = 0.
\label{eq:momcons}
\end{equation}
For instance, in the case of  viscous compressible fluids, contracting $k$ with
(\ref{mom4v}) gives
\begin{equation}
\partial_t ((k,\rho \nu) \otimes \mu) + d \left[(k,\rho \nu) \otimes  \star \nu + p k\ip \mu  - \varsigma  (k,  \star_2 \, \lie_u g)  -  \lambda (\divv u) \, k \ip \mu  \right]  = 0 .
 \label{mom5v}
\end{equation}
The density of the conserved quantity,  the $k$-directed momentum, is then $(k,\rho \nu)$ while the flux $(k,\tau)$
consists of  the terms within the square brackets.
Integrating (\ref{mom5v}) over any subregion $\Nc$ of $\Mc$ relates the time derivative of the integral of $(k,\rho\nu)$ to the transport of $(k,\rho \nu)$ across the boundary $\partial\Nc$ and the $k$-directed pressure and viscous stress on the boundary, using Stokes' theorem. In the case of $\mathbb{R}^3$ and $\mathbb{S}^2$, $(k,\rho\nu)$ corresponds to linear and angular momenta.

Second, we observe that, in the variational derivation of the equations for motion for inviscid fluids, the statement of the stationarity of the action directly gives a weak form of the equations -- with the vector field $w$ generating an arbitrary diffeomorphism regarded as a test function -- which can provide the starting point for a finite-element discretisation. The weak forms we obtain by exploiting the covariance of the action (namely (\ref{eqexp1}), (\ref{eqniceeuler}) and (\ref{eqnicealpha}) for the compressible, incompressible and Euler-$\alpha$ equations, and  (\ref{eqniceMHD}) for the additional magnetic term) are particularly simple and well suited for discretisations that preserve discrete analogues of the conserved global momenta \citep{ToHuGe,Ge}.

Third, we return to one of the motivations for using the conservation form of the equations of momentum, namely the suitability of this form when carrying out an average over fluctuations. Eulerian (Reynolds) averaging is straighforward; for the incompressible Navier--Stokes equations it leads to the 1-form valued 2-form Reynolds stress $- \overline{\nu' \otimes \star \nu'}$, where $\nu' = \nu - \overline{\nu}$ is the momentum fluctuation and the overbar denotes averaging. The situation is more complex for averages that are performed at moving rather than fixed Eulerian position, such as the thickness-weighted average used in oceanography \cite{Yo12}. The derivation of thickness-weighted average equations, leading to a geometric interpretation of the Eliassen--Palm tensor (the relevant generalisation of the Reynolds stress; see\citet{MaMa}) is the subject of ongoing work \citep{GiVa22}.

%

\appendix

\section{Computations in spherical geometry} \label{app:sphere}

We consider the 1-form valued stress $\tau$ on the sphere $\mathbb{S}^2$. In terms of  the polar and azimuthal angles $\theta$ and $\varphi$, the standard metric and associated volume (in fact area) form read
\begin{equation}
g = d \theta \otimes d \theta + \sin^2 \theta \, d \varphi \otimes d \varphi
\quad \textrm{and} \quad  \mu = \sin \theta \, d \theta \wedge d \varphi.
\label{eq:sphere}
\end{equation} 
On this two dimensional manifold the stress $\tau$ becomes a 1-form valued 1-form (rather than the 1-form valued 2-forms  used earlier for three dimensions). We write it as 
\begin{equation}
\tau = \tau_{\theta\theta} \, d \theta \otimes d \theta + \tau_{\theta\varphi} \, d \theta \otimes d \varphi + \tau_{\varphi \theta} \,  d \varphi \otimes d \theta + \tau_{\varphi\varphi} \, d \varphi \otimes d \varphi.
\end{equation}
The symmetry of the stress tensor implies a relationship between its components. Using (\ref{eq:sphere}), we find that 
\begin{subequations}
\begin{align}
\star d \theta &= (d \theta)_\sharp \ip \mu = \partial_\theta \ip \mu = \sin \theta \, d \varphi, \\ 
\star d \varphi &=  (d \varphi)_\sharp \ip \mu = \frac{1}{\sin^{2} \theta}\, \partial_\varphi \ip \mu = - \frac{1}{\sin \theta} \,d \theta,
\end{align}
\end{subequations}
hence 
\begin{equation}
\star_2 \tau = \sin \theta \, \tau_{\theta\theta} \, d \theta \otimes d \varphi - \frac{1}{\sin \theta} \, \tau_{\theta\varphi} \, d \theta \otimes d \theta +  \sin \theta \, \tau_{\varphi \theta} \,  d \varphi \otimes d \varphi -  \frac{1}{\sin \theta}\,  \tau_{\varphi\varphi} \, d \varphi \otimes d \theta.
\end{equation}
The symmetry condition in the form \eqref{eqtausym} therefore implies that
\begin{equation}
\sin \theta \, \tau_{\theta \theta} = - \frac{1}{\sin \theta} \,\tau_{\varphi \varphi}.
\label{eq:symmetry}
\end{equation}

We compute the exterior covariant derivative $\cd \tau$ using (\ref{frankelstyle}). This requires the covariant derivatives of $d \theta$ and $d \varphi$. The (Levi--Civita) connection on the sphere is determined by the relations
\begin{equation}
\nabla \partial_\theta = \cot \theta \, \partial_\varphi \otimes d \varphi \quad \textrm{and} \quad  
\nabla \partial_\varphi = \cot \theta \, \partial_\phi \otimes d \theta -  \cos \theta  \sin \theta  \, \partial_\theta  \otimes d \varphi.
\end{equation}
Using that $\nabla$ applied to contractions of basis 1-forms and basis vectors vanishes, we find the counterparts
\begin{equation}
\nabla d \theta = \cos \theta  \sin \theta  \, d \varphi \otimes d \varphi, \quad \nabla d \phi = - \cot \theta\,  ( d \theta \otimes d \varphi + d \varphi \otimes d \theta).
\end{equation}
With these expressions, the computation of $\cd \tau$ from (\ref{frankelstyle}) is straightforward:
\begin{align}
\cd \tau 
&= \cos \theta \sin \theta \, \tau_{\theta \theta} \, d \varphi \otimes d \varphi \wedge d \theta + \tau_{\theta \theta, \varphi} \, d \theta \otimes d \varphi \wedge d \theta + \tau_{\theta\varphi,\theta} \, d \theta \otimes d \theta \wedge d \varphi \nonumber  \\
& \qquad - \cot \theta \left( \tau_{\varphi \theta} \, d \theta \otimes d \varphi \wedge d \theta + \tau_{\varphi \varphi} \, d \varphi \otimes d \theta \wedge d \varphi \right) \nonumber  \\
& \qquad+ \tau_{\varphi \theta,\varphi} \, d \varphi \otimes d \varphi \wedge d \theta + \tau_{\varphi\varphi,\theta} \, d \varphi \otimes d \theta  \wedge d \varphi  \nonumber \\
&= (\tau_{\theta \varphi,\theta} - \tau_{\theta \theta,\varphi} + \cot \theta \, \tau_{\varphi\theta})  \,d \theta \otimes d \theta \wedge d \varphi \nonumber \\
& \qquad + (\tau_{\varphi\varphi,\theta}-\tau_{\varphi \theta,\varphi} - \cos \theta \sin \theta \, \tau_{\theta \theta} - \cot \theta \, \tau_{\varphi \varphi}) \,d \varphi \otimes d \theta \wedge d \varphi \nonumber \\
&= (\tau_{\theta \varphi,\theta} - \tau_{\theta \theta,\varphi} + \cot \theta \, \tau_{\varphi\theta})  \,d \theta \otimes d \theta \wedge d \varphi + (\tau_{\varphi\varphi,\theta}-\tau_{\varphi \theta,\varphi}) \,d \varphi \otimes d \theta \wedge d \varphi,
\end{align}
using the symmetry property (\ref{eq:symmetry}) to simplify the penultimate line.

It is interesting to verify explicitly the property (\ref{eq:ktau}) that contraction of the first leg of $\cd \tau$ with a Killing vector field $k$ yields the (metric-independent) pairing $(k,\tau)$. The sphere $\mathbb{S}^2$ has the three Killing fields
\begin{equation}
k_1 = - \sin \varphi \, \partial_\theta  - \cot \theta \cos \varphi \, \partial_\phi,  \quad 
k_2 = \cos \varphi \, \partial_\theta - \cot \theta \sin \varphi \, \partial_\phi \quad
\textrm{and} \quad k_3 = \partial_\phi,
\end{equation}
corresponding to rotation about the $x$, $y$ and $z$ axes. We have
\begin{subequations}
\begin{align}
(k_1,\cd \tau) &= \left[ -\sin \varphi \left(\tau_{\theta \varphi,\theta} - \tau_{\theta \theta, \varphi} + \cot \theta \, \tau_{\varphi\theta} \right) - \cot \theta \cos \varphi \left(\tau_{\varphi\varphi,\theta}-\tau_{\varphi \theta,\varphi}\right) \right]  d \theta \wedge d \varphi,  \\
(k_2,\cd \tau) &= \left[ \cos \varphi \left(\tau_{\theta \varphi,\theta} - \tau_{\theta \theta, \varphi} + \cot \theta \, \tau_{\varphi\theta} \right) - \cot \theta \sin \varphi \left(\tau_{\varphi\varphi,\theta}-\tau_{\varphi \theta,\varphi}\right) \right]  d \theta \wedge d \varphi,  \\
(k_3,\cd \tau) &= \left(\tau_{\varphi\varphi,\theta}-\tau_{\varphi \theta,\varphi}\right)  d \theta \wedge d \varphi, 
\end{align}
\end{subequations}
while 
\begin{subequations}
\begin{align}
(k_1, \tau) &=   \left( - \sin \varphi\,  \tau_{\theta\theta} -  \cot \theta \cos \varphi \, \tau_{\varphi\theta} \right) d\theta +  \left( - \sin \varphi\,  \tau_{\theta\varphi} -  \cot \theta \cos \varphi \, \tau_{\varphi\varphi} \right) d\varphi, \\
(k_2, \tau) &=     \left(\cos \varphi\,  \tau_{\theta\theta} -  \cot \theta \sin \varphi \, \tau_{\varphi\theta} \right) d\theta +  \left( \cos \varphi\,  \tau_{\theta\varphi} -  \cot \theta \sin \varphi \, \tau_{\varphi\varphi} \right) d\varphi, \\
(k_3, \tau) &= \tau_{\varphi\theta} \, d\theta + \tau_{\varphi\varphi} \, d\varphi. 
\end{align}
\end{subequations}
A direct computation using (\ref{eq:symmetry}) gives $(k_i, \cd\tau) = d(k_i,\tau)$ for $i=1,\,2,\,3$, as expected from (\ref{eq:ktau}). This implies conservation laws of the form (\ref{eq:momcons}) for the angular momenta $(k_i,\rho \nu \otimes \mu)$, explicitly 
\begin{subequations}
\begin{multline}
\partial_t ( \sin \theta \sin \varphi \, \rho \nu_\theta + \cos \theta \cos \varphi \, \rho \nu_\varphi) + \partial_\theta \left(\sin \varphi \, \tau_{\theta\varphi} + \cot \theta \cos \varphi \, \tau_{\varphi\varphi} \right) \\ - \partial_\varphi \left( \sin \varphi \, \tau_{\theta \theta} + \cot \theta \cos \varphi \, \tau_{\varphi\theta} \right) = 0,
\end{multline}
\begin{multline}
\partial_t ( \sin \theta \cos \varphi \, \rho \nu_\theta - \cos \theta \sin \varphi \, \rho \nu_\varphi) + \partial_\theta \left( \cos \varphi \, \tau_{\theta\varphi} - \cot \theta \sin \varphi \, \tau_{\varphi\varphi} \right) \\ - \partial_\varphi \left(  \cos \varphi \, \tau_{\theta \theta} - \cot \theta \sin \varphi \, \tau_{\varphi\theta} \right) = 0,
\end{multline}
\begin{equation}
\partial_t ( \sin \theta \, \rho \nu_\varphi) + \partial_\theta  \tau_{\varphi\varphi}   - \partial_\varphi \tau_{\varphi\theta}  = 0.
\end{equation}
\end{subequations}

\section{Variational derivation of (\ref{eulercomp})} \label{app:EP}

We detail the variational derivation of the Euler equations in (\ref{eulercomp}) from the action (\ref{eqaction}). Starting with condition (\ref{eqvary}) for the stationarity of the action, we use integration by parts to rewrite each term as a pairing with the undifferentiated $w$. The first term is given in (\ref{uw_t}); the others are
\begin{subequations}
 \label{eqbyparts}
\begin{align}
 & g(u,  \lie_w u)\,  \rho \mu  =  (- \lie_u w, \nu \otimes \rho \mu) \byparts  (w, \lie_u (\rho\nu\otimes  \mu) ),
 \label{eqlots2}\\
  & \tfrac{1}{2} g(u,u)\,  \lie_w (\rho \mu)  \byparts - \rho  \mu \, \lie_w  \tfrac{1}{2} g(u,u) = - (w, \tfrac{1}{2}\rho\,  d g(u,u)\otimes  \mu ) ,\\
  & (\rho e)_\rho \, \lie_w (\rho \mu)  \byparts -  \rho  \mu \, \lie_w  [(\rho e)_\rho]  = -  (w,   \rho \, d (\rho e)_\rho  \otimes  \mu )  , \label{eqaa} \\
&  \rho e_s  \, (\lie_w s )\,  \mu   = \rho e_s  \, ( w, ds) \, \mu = (w,   \rho e_s \, ds \otimes \mu) , \label{eqaaa}
\end{align}
\end{subequations}
on using that, for any scalar field $f$, $\lie_w(f \mu) = d(f w \ip \mu) \byparts 0$ by Cartan's formula. To explain, as an example, one of these in more detail, consider (\ref{eqlots2}). We write first 
\begin{equation}
 ( \lie_u w, \nu \otimes \rho \mu)  = \lie_u [ (w\ip\nu) \, \rho \mu] - w \ip \lie_u (\nu \otimes \rho \mu) . 
\end{equation}
We have from Cartan's formula (\ref{eqcartan}) applied to the term we wish to remove,  $ \lie_u [ (w\ip\nu) \, \rho \mu] = d [  (w\ip\nu) \, u \ip \rho \mu]$, and then on integrating over $\Mc$ we find 
\begin{equation}
\int_\Mc d [(w\ip\nu) \, u \ip \rho \mu ] = \int_{\partial \Mc} (w\ip\nu) \, u \ip \rho \mu  = 0 , 
\end{equation}
using (\ref{eqstokestheorem}) and the boundary condition that $u \parallel \partial \Mc$: if a surface element is defined by vectors $a$ and $b$ at a point, then $u \ip \mu(a,b) = \mu(u,a,b)$ vanishes as $u$ is contained in the vector space spanned by $a$ and $b$. 

Introducing the various formulae (\ref{eqbyparts}) into (\ref{eqvary}) gives
\begin{equation}
\int dt \int_{\Mc}  \left[ - (w,(\partial_t + \lie_u) ( \rho \nu \otimes \mu) )   + (w,   \tfrac{1}{2}\rho \,d g(u,u) \otimes  \mu) {-}
(w, [  \rho \,d  (\rho e)_\rho - \rho e_s \, ds]  \otimes \mu)  \right]  = 0 . 
\label{eqvary1}
\end{equation}
We use the thermodynamic definitions that  $T=\partial_s e$ is the temperature and $h=(\rho e)_\rho = e + p / \rho$  is the enthalpy, together with $\d h =  \rho^{-1} \,\d p + T\, \d s$ to simplify the last terms. Requiring this integral to be zero for arbitrary $w$ recovers the equation of motion as precisely (\ref{eqeuler1}).

%

\section{Shallow water equations in conservation form} \label{apshallow}

In this appendix, we derive conservation forms for the shallow water and  MHD shallow water models. We consider a two-dimensional manifold $\Mc$ supporting a (two-dimensional) fluid flow $u$ and scalar height field $h$; flows and magnetic fields are taken parallel to any boundary of $\Mc$. 
The shallow water action is given by 
\begin{equation}
\Ac[\phi] = \int dt \int_{\Mc} (  \tfrac{1}{2} h g(u, u)  - \tfrac{1}{2} h^2 ) \, \mu,
\label{eqactionSW} 
\end{equation}
where the height field transport is governed by conservation of mass,
\begin{equation}
(\partial_t +  \lie_u ) (h\mu)  = 0, 
\label{eqconsmassSW}
\end{equation}
or equivalently $ h \mu     = \phi_* (h_0 \mu)$, where $h_0$ is the initial height.  When the flow map is varied we have
\begin{equation}
\dt{}{\eps} \Big|_{\eps=0} (h_\eps \mu) = - \lie_w (h\mu) = - \divv (h {w} ) \, \mu. 
\label{eqhepsSW}
\end{equation}
Varying the action (\ref{eqactionSW}) gives 
\begin{equation}
 \left. \dt{}{\eps} \right|_{\eps=0}\Ac [ \psi_\eps\circ\phi] = \int dt \int_{\Mc}[h g(u , \partial_t w + \lie_u w)  \, \mu  - (  \tfrac{1}{2} g(u, u) - h )\,  \lie_w (h\mu) ]  = 0,
\label{eqvaryactionSW} 
\end{equation}
and so we gain 
\begin{equation}
\partial_t (  h \nu \otimes \mu) + \lie_u (h\nu \otimes \mu) + d ( -  \tfrac{1}{2} g(u, u) + h   ) \otimes h \mu = 0.
\end{equation}
Given (\ref{eqconsmassSW}) we can write this equation in the usual form 
\begin{equation}
\partial_t  \nu + \lie_u \nu   -  \tfrac{1}{2} dg(u, u)  + d h = 0.
\end{equation}

If on the other hand we apply the covariance of the action (\ref{eqlieaction}), we have 
\begin{equation}
 \int dt \int_{\Mc} [  \tfrac{1}{2} (\lie_w g)(u,u)\,  h \mu + g(u,\lie_w u)\,  h \mu + ( \tfrac{1}{2} g(u,u) -h) ( \lie_w h) \, \mu +
  ( \tfrac{1}{2} h g(u,u) -\tfrac{1}{2}h^2 ) ( \lie_w \mu)   ]  = 0. 
\label{eqactionSWcov} 
\end{equation}
Combining with  (\ref{eqvaryactionSW}) and tidying gives 
\begin{equation}
 \int dt \int_{\Mc} [ \tfrac{1}{2} (\lie_w g) (u,u) \, h \mu +  g (u, \partial_t w) \, h \mu  + \tfrac{1}{2} h^2 \, \lie_w \mu ]  = 0 , 
\label{eqactionSWcombine} 
\end{equation}
with the conservation form easily derived as 
\begin{equation}
\partial_t (h\nu \otimes \mu) + \cd ( h\nu \otimes \star\nu +\tfrac{1}{2} h^2 \mu ) = 0 .
\end{equation}

Magnetic fields can also be incorporated into shallow water systems and the resulting modelling is relevant to the Solar tachocline and other stratified MHD systems in astrophysics  \citep{Gi00,De02}. In our setting, given any two points $x$ and $y$ of our two-dimensional $\Mc$, what is key is the magnetic flux between these points and so we define a scalar magnetic potential $a$ (up to a constant) so that this flux is $a(y)-a(x)$. Since these points, i.e.\ these columns of fluid in the real system, move as Lagrangian markers in the flow, the flux between them is conserved and so $a$ evolves according to 
\begin{equation}
(\partial_t  + \lie_u )a = 0 . 
\end{equation}
We then set $h\beta = da$ where the magnetic flux $\beta$ is now a 1-form such that the total flux through a 1-dimensional surface element in $M$, that is integrated over the fluid layer from base to $h$, is given by $h \beta$. This satisfies $d(h\beta) =0$ and also 
\begin{align}
&  (\partial_t + \lie_u) (h \beta)  = 0 . 
\label{eqliehB}
\end{align}
The corresponding magnetic vector field $b$ is related to $\beta$ through $b\ip \mu = \beta$ or, equivalently $\star \beta = b_\flat$. It satisfies $\divv (hb)=0$ and, from (\ref{eqconsmassSW}) and (\ref{eqliehB}), 
\begin{equation}
(\partial_t + \lie_u) b = 0 . 
\end{equation}
Note that there is no $ b \divv u$ term present, in contrast to (\ref{eqbtensorweight}): the effects of non-zero divergence of the flow $u$ are absorbed into the height field $h$.

The action is $\Ac-\Bc$, with $\Ac$ the shallow-water action (\ref{eqactionSW}) and $\Bc$ the magnetic term
\begin{equation}
\Bc[\phi] = \int dt \int_{\Mc}  \tfrac{1}{2} h g(b , b ) \, \mu .
\end{equation}
When the path is varied we have (\ref{eqhepsSW}) and 
\begin{align} 
\left. \dt{}{\eps} \right|_{\eps=0} b_\eps &=  - \lie_w b    , 
\label{eqtransbsw} 
\end{align}
(contrast (\ref{eqtransb})). Hence we find that
\begin{equation}
 \left. \dt{}{\eps} \right|_{\eps=0}\Bc [ \psi_\eps\circ\phi] = \int dt \int_{\Mc} [ - g(b, \lie_w b) \, h \mu -  \tfrac{1}{2} g(b,b) \, \lie_w (h\mu) ]   . 
\label{eqvaryactionSWMHD} 
\end{equation}
Integrating by parts and using the arbitrariness of $w$ we obtain the equation of motion 
\begin{align}
\partial_t (  h \nu \otimes \mu) & + \lie_u (h\nu \otimes \mu) + d ( h -  \tfrac{1}{2} g(u, u)  ) \otimes h \mu  = 
  \lie_b( h {\star} \beta \otimes \mu) 
-  \tfrac{1}{2} dg(b,b) \otimes h\mu  .
\end{align}
Using (\ref{eqconsmassSW}) and noting that $\lie_b (h \mu)  = d ( b \ip h\mu) = \mu \divv (hb) = 0 $, we can write this as 
\begin{align}
\partial_t  \nu & + \lie_u \nu  + d ( h -  \tfrac{1}{2} g(u, u)  ) =   \lie_b {\star} \beta -  \tfrac{1}{2} dg(b,b)  . 
\end{align}

If instead we apply the covariance (\ref{eqlieaction}) the  terms associated with $\Bc$ are 
\begin{equation}
 \int dt \int_{\Mc} [  \tfrac{1}{2} (\lie_w g) (b,b) \, h \mu + g(b, \lie_w b) \, h\mu +  \tfrac{1}{2} g(b,b) \, \lie_w (h \mu)   ].  
\label{eqactionSWMHD} 
\end{equation}
Combining this with the path variation (\ref{eqvaryactionSWMHD}) leaves only 
\begin{equation}
  \left. \dt{}{\eps} \right|_{\eps=0}\Bc [ \psi_\eps\circ\phi] = \int dt \int_{\Mc}   \tfrac{1}{2} (\lie_w g) (b,b) \, h \mu , 
\label{eqactionSWMHDcombine} 
\end{equation}
giving the conservation version of shallow water MHD as 
\begin{equation}
\partial_t (h\nu \otimes \mu) + \cd ( h\nu \otimes   \star \nu + \tfrac{1}{2} h^2 \mu ) = \cd ( h {\star} \beta \otimes  \beta )  .
\end{equation}
Note that there is no magnetic pressure term here, that is the term $ - \tfrac{1}{2} d g(b,b)\mu$ present in (\ref{eqmhdcons}). Although shallow water dynamics has many attributes of compressible fluid flow, with the height field $h$ playing the role of pressure, the underlying fluid dynamics is incompressible and the magnetic pressure does not emerge in the resulting equations \citep{Gi00,De02}.

\medskip
\noindent
\textbf{Acknowledgements}

\noindent
{ADG is grateful to the Leverhulme Trust, who supported this work through the award of a Research Fellowship (grant RF-2018-023), and to EPSRC for support under the research grant EP/T023139/1. The authors thank Darryl Holm, Alexander Morozov, Marcel Oliver and Andrew Soward for helpful discussions relating to geometrical fluid dynamics,  MHD and viscoelastic fluids.}

\medskip
\noindent
\textbf{Data Access Statement}

\noindent
No data were created or analysed in this study.



\vskip2pc



\begin{thebibliography}{9}

\bibitem[Arnold (1966)]{Ar66}
Arnold, V. I. 1966
Sur la g{\'e}om{\'e}trie diff\'erentielle des groupes de {L}ie de dimension infinie et ses applications {\`a}
l'hydrodynamique des fluides parfaits.
\textit{Ann. Inst. Fourier} \textbf{16}, 316--361.

\bibitem[Arnold and Khesin(1998)]{ArKh98}
Arnold, V. I.  Khesin, B. A. 1998
\textit{Topological methods in hydrodynamics}. 
Springer--Verlag, Applied Mathematical Sciences, vol.\ 125. 

\bibitem[Batchelor(1967)]{Ba}
Batchelor, G. K. 1967
\textit{An introduction to fluid dynamics.} 
Cambridge University Press. 

\bibitem[Besse and Frisch(2017)]{BeFr}
Besse, N. and Frisch, U. 2017 
Geometric formulation of the Cauchy invariants for incompressible Euler flow in flat and curved spaces.
\textit{J. Fluid Mech.} \textbf{825}, 412--478.

\bibitem[Bird et al.(1977)]{BiHaArCu}
Bird, R. B., Hassager, O., Armstrong, R. C. and Curtiss, C. F. 1977 
\textit{Dynamics of polymeric liquids, vol.\ 2: kinetic theory}.
Wiley.

\bibitem[Dellar(2002)]{De02}
Dellar, P. J. 2002
Hamiltonian and symmetric hyperbolic structures of shallow water magnetohydrodynamics.
\emph{Phys. Plasmas} {\bf 9}, 1130--1136.

\bibitem[Frankel(1997)]{Fr}
Frankel T. 1997
\textit{The geometry of physics}. Cambridge University Press. 

\bibitem[Gay-Balmaz and  Ratiu(2005)]{GaRa}
Gay-Balmaz, F. and  Ratiu, T. S. 2005
The Lie-Poisson structure of the LAE-$\alpha$ equation.
\emph{Dyn. PDEs} {\bf 2}, 25--57.

\bibitem[Gerritsma(2014)]{Ge}
Gerritsma, M. 2014
Structure-preserving discretization for continuum models. In \textit{Proc. 21st International Symposium on Mathematical Theory of Systems and Networks}, July 7-11, 2014. Groningen, The Netherlands.

\bibitem[Gilbert, Riedinger and Thuburn(2014)]{GiRiTh}
Gilbert, A. D., Riedinger, X. and Thuburn, J. 2014
On the form of the viscous term for two dimensional Navier--Stokes flows.
\textit{QJMAM} \textbf{67}, 205--228. 

\bibitem[Gilbert and Vanneste(2018)]{GiVa} 
Gilbert,  A. D. and Vanneste, J. 2018 
Geometric generalised Lagrangian mean theories.
\textit{J. Fluid Mech.} \textbf{839}, 95--134.

\bibitem[Gilbert and Vanneste(2021)]{GiVa19b}
Gilbert, A. D. and Vanneste, J. 2019
A geometric look at MHD and the Braginsky dynamo,
\textit{Geophys. Astrophys. Fluid Dynam.}, {\bf 115}, 436--471. 

\bibitem[Gilbert and Vanneste(2022)]{GiVa22}
Gilbert, A. D. and Vanneste, J. 2022
A geometric look at thickness-weighted averaging.
In preparation. 

\bibitem[Gilman(2000)]{Gi00}
Gilman, P. A. 2000
Magnetohydrodynamic `shallow water' equations for the solar tachocline.
\emph{Astrophys. J.} {\bf 544}, L79--82.

\bibitem[Gotay and Marsden(1992)]{GoMa}
Gotay M. J. and Marsden J. E. 1992
Stress--energy--momentum tensors and the Belinfante-Rosenfield formula. 
In: \textit{Mathematical aspects of classical field theory. Contemp. Math.}, No.\ 132. American Mathematical Society, 367--392.

\bibitem[Hawking and Ellis(1973)]{HaEl}
Hawking, S. W. and Ellis, G. F. R. 1973
\textit{The large scale structure of space-time}.
Cambridge University Press. 

\bibitem[Holm(1999)]{Ho99}
Holm, D. D. 1999 
Fluctuation effects on 3D Lagrangian mean and Eulerian mean fluid motion.
\textit{Physica D}  \textbf{133}, 215--269.

\bibitem[Holm(2002)]{Ho02a}
Holm, D. D. 2002
Variational principles for Lagrangian-averaged fluid dynamics.
\textit{J. Phys. A. Math. Gen.} \textbf{35}, 679--668.


\bibitem[Holm, Marsden and Ratiu(1998)]{HoMaRa}
Holm, D. D., Marsden, J. E. and Ratiu, T. 1998
The Euler--Poincar\'e equations and semi-direct products with applications to continuum theories.
\textit{Adv. Math.} \textbf{137}, 1--81.

\bibitem[Holm, Schmah and Stoica(2009)]{HoScSt09}
Holm, D. D., Schmah, T. and Stoica, C. 2009
\textit{Geometric mechanics and symmetry: from finite to infinite dimensions}.
Oxford University Press.

\bibitem[Kanso et al.(2007)]{KaArToYaMaDe}
Kanso, E., Arroyo, M., Tong, Y., Yavari, A., Marsden, J. E. and Desbrun, M. 2007 
On the geometric character of stress in continuum mechanics.
\textit{Z. angew. Math. Phys.} \textbf{58}, 843--856. 

\bibitem[Lindborg and Nordmark(2022)]{LiNo}
Lindborg, E. and Nordmark, A. 2022
Two-dimensional turbulence on a sphere.
\textit{J. Fluid Mech.} \textbf{933}, A60.

\bibitem[Maddison and Marshall(2013)]{MaMa}
Maddison, J. R. and Marshall, D. P. 2013
The Eliassen--Palm flux tensor.
\textit{J. Fluid Mech.} \textbf{729}, 62--102.

\bibitem[Marsden and Hughes (1983)]{MaHu}
Marsden, J. E. and Hughes, T. J. R. 1983 
\textit{Mathematical foundations of elasticity.} Dover.

\bibitem[Marsden, Ratiu and Shkoller(2000)]{MaRaSh}
Marsden, J. E., Ratiu, T. S. and Shkoller, S. 2000
The geometry and analysis of the averaged Euler
equations and a new diffeomorphism group.
\emph{Geom. Funct. Anal.} {\bf 10}, 582--599.

\bibitem[Marsden and Shkoller(2003)]{MaSh}
Marsden, J. E and Shkoller, S. 2003 
The anisotropic Lagrangian averaged Euler and Navier--Stokes equations.
\textit{Arch. Ration. Mech. Anal.} \textbf{166}, 27--46.

\bibitem[Morozov and Spagnolie(2015)]{MoSp}
Morozov, A. and Spagnolie, S. E. 2015
Introduction to complex fluids. In 
\textit{Complex fluids in biological systems}, S. E. Spagnolie (ed.), Springer.

\bibitem[Morrison(1998)]{Mo98}
Morrison, P. J. 1998
Hamiltonian description of the ideal fluid.
\textit{Rev. Mod. Phys.}, \textbf{70}, 467--521.

\bibitem[Newcomb(1962)]{Ne62}
Newcomb, W. A. 1962 Lagrangian and Hamiltonian methods in magnetohydrodynamics. 
\textit{Nucl. Fusion Suppl.}, part 2, 451--463.

\bibitem[Oldroyd(1950)]{Ol50}
Oldroyd, J. G.  1950 On the formulation of rheological equations of state.
\textit{Proc. Roy. Soc. Lond. A} {\bf 200}, 523--541.

\bibitem[Oliver(2017)]{Ol}
Oliver, M. 2017
Lagrangian averaging with geodesic mean.
\textit{Proc. R. Soc. Lond. A} \textbf{473}, 20170558. 

\bibitem[Oliver and Vasylkevych(2019)]{OlVa19}
Oliver, M. and Vasylkevych, S. 2019
Geodesic motion on groups of diffeomorphisms with $H^1$ metric as geometric generalised Lagrangian mean theory.
\emph{Geophys. Astrophys. Fluid Dynam.} {\bf 113}, 466--490. 

\bibitem[Roberts and Soward(2006)]{RoSo06b}
Roberts, P. H. and Soward, A. M. 2006
Covariant description of non-relativistic magnetohydrodynamics.
\textit{Geophys. Astrophys. Fluid Dynam.} {\bf 100}, 485--502.

\bibitem[Salmon(1988)]{Sa88}
Salmon, R. 1988
Hamiltonian fluid mechanics.
\textit{Ann. Rev. Fluid Mech.}, \textbf{20}, 225--256.

\bibitem[Schutz(1980)]{Sc80}
Schutz, B. 1980 
\textit{Geometrical methods of mathematical physics}. 
Cambridge University Press.

\bibitem[Shkoller(1998)]{Sh98}
Shkoller, S. 1998
Geometry and curvature of diffeomorphism groups with H1 metric and mean hydrodynamics.
\emph{J. Funct. Anal.} {\bf 160}, 337--365.

\bibitem[Shkoller(2000)]{Sh00}
Shkoller, S. 2000 
Analysis on groups of diffeomorphisms of manifolds with boundary and the averaged motion of a fluid.
\emph{J. Diff. Geom.} {\bf 55}, 145--191.

\bibitem[Toshniwal, Huijsmans and Gerritsma(2014)]{ToHuGe}
Toshniwal, D., Huijsmans, R. H. M. and Gerritsma, M. I. 2014
A geometric approach towards momentum conservation. In: Aza\"\i ez, M., El Fekih, H., Hesthaven, J. (eds.) 
\textit{Spectral and high order methods for partial differential equations},  Springer.

\bibitem[Webb(2018)]{We18}
Webb, G. 2018 
\textit{Magnetohydrodynamics and fluid dynamics: action principles and conservation laws}. Springer.

\bibitem[Young(2012)]{Yo12}
Young, W. R. 2012
An exact thickness-weighted average formulation of the Boussinesq equations.
\textit{J. Phys. Oceanogr.} \textbf{42}, 692--707.



\end{thebibliography}
\end{document}